\documentclass[lettersize,journal]{IEEEtran}
\usepackage{amsmath,amsfonts}
\usepackage{algorithmic}
\usepackage{algorithm}
\usepackage{amssymb}
\usepackage{array}
\usepackage{textcomp}
\usepackage{stfloats}
\usepackage{url}
\usepackage{verbatim}
\usepackage{graphicx}
\usepackage{cite}
\usepackage{color}
\usepackage{mathtools}
\usepackage{float}
\usepackage{caption}
\usepackage{xcolor}
\usepackage{svg}
\usepackage{mathtools}
\usepackage{subcaption}
\begin{document}
		\title{Pinching-Antenna Systems-Assisted SWIPT: A Rate-Energy Trade-off Perspective}
				\author{Qi~Yang,
					Kai~Liu,
					Jingjing Zhao,~\IEEEmembership{Senior Member,~IEEE,}
					Kaiquan~Cai,
					Xidong~Mu, \\
					and Yuanwei~Liu,~\IEEEmembership{Fellow,~IEEE}
					\thanks{Q. Yang, K. Liu, J. Zhao, K. Cai are with the School of Electronics and Information Engineering, Beihang University, 100191, Beijing, China, and also with the State Key Laboratory of CNS/ATM, 100191, Beijing, China. (e-mail:\{yangqi01, liuk, jingjingzhao, caikq\}@buaa.edu.cn). }
						
					\thanks{X. Mu is with the Centre for Wireless Innovation (CWI), Queen's University Belfast, Belfast, BT3 9DT, U.K. (e-mail: x.mu@qub.ac.uk). }

					\thanks{Y. Liu is with the Department of Electrical and Electronic Engineering, the University of Hong Kong, Hong Kong, China (e-mail: yuanwei@hku.hk). }}
	
		\maketitle
	
	\begin{abstract}
	This paper investigates the rate-energy trade-off for pinching-antenna systems (PASS)-assisted simultaneous wireless information and power transfer (SWIPT) systems.
	Both the single information user (IU)/energy user (EU) and multiple IUs/EUs scenarios are considered.
	1) For the single IU/EU scenario, a pinching beamforming optimization problem is formulated for simultaneously maximizing data rate and harvested energy.
	To tackle this problem, a two-stage algorithm is proposed.
	Specifically, the successive convex approximation (SCA) method is first invoked for minimizing the large-scale path loss, which is followed by the fine-tuning method for the phase alignment.
	2) For the multiple IUs/EUs scenario, three multiple access schemes are considered, i.e., frequency division multiple access (FDMA), time division multiple access (TDMA), and non-orthogonal multiple access (NOMA).
	The corresponding multi-objective optimization problem (MOOP) that simultaneously maximizes the minimum data rate and minimum harvested energy is formulated for ensuring users' fairness.
	To address this problem, we adopt the $\epsilon$-constraint method to first convert the intractable MOOPs to single-objective optimization problems (SOOPs). 
	Then, for the SOOP under each multiple access protocol, the particle swarm optimization (PSO) and convex optimization methods are adopted for solving the pinching beamforming and resource allocation problems, respectively.
	Simulation results unveil that: 
	i) PASS can achieve a significantly superior rate-energy region compared to conventional fixed-position antenna systems for pinching beamforming; 
	and ii) by exploiting the time-switching feature, TDMA can outperform both NOMA and FDMA for the multiple IUs/EUs scenario.
	\end{abstract}
		
	
	\section{Introduction}

	\IEEEPARstart{W}{ith} the rapid development of the internet of things (IoT), ensuring seamless communication and energy sustainability for a multitude of devices has become a critical challenge for wireless networks~\cite{KuAdvances2016, WangUltra2022, LiuEfficient2022}.
	Conventional energy supply methods, such as battery replacement and wired connections, are severely constrained by high maintenance costs, limited deployment flexibility, and adverse environmental impacts~\cite{CaiBattery2023}.
	To overcome these constraints, simultaneous wireless information and power transfer (SWIPT) has emerged as a promising solution.
	By utilizing the same radio frequency (RF) signals, SWIPT enables simultaneous wireless information transfer (WIT) for information users (IUs) and wireless power transfer (WPT) for energy users (EUs)~\cite{ZhangMIMO2013}.
	However, the practical deployment of SWIPT is fundamentally limited by the quality of the wireless channel. 
	Poor channel conditions result in a limited channel gain, which degrades both reliable information transmission for IUs and efficient energy harvesting for EUs, even rendering the latter impractical.
	Therefore, establishing robust and high-gain communication links is the critical problem for advancing practical SWIPT systems.
	
	A flexible solution for improving the SWIPT performance is presented by multiple-input multiple-output (MIMO) technology, which exploits spatial degrees of freedom (DoFs) through beamforming design~\cite{WangTutorial2024}.
	Specifically, MIMO-assisted SWIPT systems employ beamforming to steer communication/energy signals toward desired IUs/EUs, which reduces interference in unintended directions and enhances the overall performance of SWIPT systems~\cite{FemeniasSWIPT2021}.
	However, the fixed antenna positions restrict the full utilization of spatial resources in MIMO systems, thereby constraining their spatial multiplexing performance.
	To overcome the inherent limitations of conventional MIMO, various flexible-antenna technologies, such as movable antenna (MA)~\cite{XiaoChannel2024} and fluid antenna (FA)~\cite{NewTutorial2025}, have been recently proposed.
	By moving the positions of antennas within a confined range, MAs and FAs can actively reconfigure the wireless propagation environment, thereby significantly enhancing the effective channel gain of SWIPT systems~\cite{ZhouFluid2025, SkouroumounisFA2025, DongMovable2025, HuangWeighted2025}.
	However, MAs and FAs are restricted to move within a few wavelengths, possessing inadequate capability for overcoming free-space path loss.
	Moreover, when the communication link between the base station (BS) and the user is blocked, the antenna movement within a limited region is generally insufficient to establish a line-of-sight (LoS) link \cite{MaMovable2025, XiaoMultiuser2024}.

	As a novel paradigm in flexible-antenna technologies, pinching-antenna systems (PASS) have been viewed as a promising solution to combat the large-scale path loss. 
	Specifically, PASS utilize a dielectric waveguide as its transmission medium, along which pinching-antennas (PAs) made of low-cost dielectric materials can be positioned at any desired positions for signal transmission~\cite{liu2025pinching, TyrovolasPerformance2025, XuToward2025}.
	This unique architecture enables PAs to be deployed at locations much closer to users, thereby effectively establishing robust LoS links and obtaining scalable pinching beamforming gain. 
	Thanks to the superiority of PASS for enhancing the channel qualities, the integration of PASS with SWIPT can potentially improve the performance of both information transmission and energy harvesting, which represents a promising research direction~\cite{YangPinching2025}.
	
	\vspace{-10pt}
	
	\subsection{Related Works}
	Some early efforts~\cite{suzuki2022pinching, OuyangArray2025, DingFlexible2025, LvBeam2025, XieLow2025, WangAntenna2025, XuRate2025, TegosMinimum2025, EnergyMing2025} have been made to deploy PASS in wireless communication systems. 
	NTT DOCOMO first demonstrated the PASS technology in 2022, which gains increasing attention for its potential to revolutionize wireless communication~\cite{suzuki2022pinching}.
	Then, the authors of~\cite{OuyangArray2025} theoretically analyzed the optimal number of PAs to maximize the array gain.
	From a performance-limit perspective, an approximate solution for the ergodic sum rate upper bound of PASS was derived in~\cite{DingFlexible2025} for orthogonal multiple access (OMA) and non-orthogonal multiple access (NOMA). 
	Moreover, the authors of~\cite{LvBeam2025} proposed a scalable codebook design for PASS beam training, significantly reducing the corresponding training overhead.
	A significant body of subsequent work has focused on designing effective optimization algorithms for determining the PA placement along the waveguide. 
	In~\cite{XieLow2025}, a PA positioning algorithm with the aim of  path-loss minimization was designed for both time division multiple access (TDMA) and NOMA scenarios. 
	The authors of~\cite{WangAntenna2025} proposed a matching theory-based PAs activation method to maximize the sum rate in NOMA-assisted downlink PASS. 
	For the single-user case, a two-stage algorithm was proposed in~\cite{XuRate2025} to optimize the positions of PAs, thus maximizing the downlink transmission rate.
	In~\cite{TegosMinimum2025}, the authors addressed the problem of maximizing the minimum uplink data rate by iteratively optimizing the pinching beamforming and resource allocation.
	Furthermore, the authors of~\cite{EnergyMing2025} maximized the energy efficiency by jointly optimizing the power allocation and pinching beamforming in the NOMA-assisted uplink PASS.
	
	As a further advancement, the application of PASS has been extended beyond communication to physical layer security~(PLS)~\cite{ZhongPhysical2025}, wireless sensing~\cite{WangWireless2025}, and integrated sensing and communications~(ISAC)~\cite{ZhangIntegrated2025}.
	Specifically, the authors of~\cite{ZhongPhysical2025} proposed a PASS-assisted PLS framework for enhancing the channel gain to the intended user by optimizing pinching beamforming.
	To improve the multi-target sensing accuracy, a \text{Cram\'er-Rao} bound minimization problem was solved in~\cite{WangWireless2025} by jointly optimizing transmit waveform and the positions of PAs.
	Moreover, the authors of~\cite{ZhangIntegrated2025} investigated a separated PASS-assisted ISAC design, where a penalty-based alternating optimization (AO) algorithm was adopted to optimize pinching beamforming, thus enhancing the sensing performance.
	Recent studies~\cite{SpatiallyRuihong2025, OnZhang2025} have begun to investigate the performance of PASS-assisted SWIPT systems.
	For the fundamental scenario of a single PA serving a power-splitting-based SWIPT user, the work in~\cite{SpatiallyRuihong2025} aimed at maximizing the received signal-to-noise ratio (SNR) by jointly optimizing the power allocation and PA position.
	Furthermore, the authors of~\cite{OnZhang2025} proposed three waveguide deployment schemes for a single PA-assisted SWIPT system under the hybrid time switching and power splitting protocol, and derived a closed-form expression for achievable rate and harvested energy under each scheme.
	
	\subsection{Motivations and Contributions}
	
	Although recent works~\cite{SpatiallyRuihong2025, OnZhang2025} have investigated the PASS-assisted SWIPT system, the fundamental rate-energy region brought by the pinching beamforming remains unexplored.
	The primary challenge in characterizing this region stems from the inherent trade-off between the objectives of IUs and EUs.
	Even in a simple single IU/EU scenario,  both users’ performance is jointly influenced by the shared configuration of PAs positions, which means that optimizing PAs positions for improving the performance of IUs may potentially degrade the performance of EUs, and vice versa.
	Hence, the optimization of PAs positions should consider both requirements for the IU and EU by exploring the complete rate-energy region.
	This challenge becomes more complex in the multiple IUs/EUs case, where the multi-user resource allocation is intrinsically coupled with PAs positions optimization, yielding a more challenging investigation on the communication rate and energy harvesting trade-off.
	
	To address the above issues, in this work, we consider both the PASS-assisted single IU/EU and multiple IUs/EUs scenarios, and investigate the pinching beamforming problem with the aim of maximizing both the communication and power transfer performance.
	The main contributions of this paper are summarized as follows.
	
	\begin{itemize}
		\item 
		We propose a PASS-assisted SWIPT system, where multiple PAs are deployed to serve both IUs and EUs, aiming to explore the trade-off between data rate and harvested energy. 
		Both single IU/EU and multiple IUs/EUs scenarios are investigated.
		\item 
		For the single IU/EU case, we formulate a pinching beamforming optimization problem to maximize the weighted sum of the channel gains for both users.
		To explore the effective rate-energy region, we propose a two-stage pinching beamforming algorithm. 
		The successive convex approximation (SCA) method is first invoked to obtain coarse PAs positions for combating large-scale path loss.
		Then, PAs positions are further fine-tuned to achieve constructive combinations for both signals received at IU and EU.
		\item 
		For the multiple IUs/EUs case, we formulate multi-objective optimization problems (MOOPs) that maximize the minimum data rate for IU and the minimum harvested energy for EU under three multiple access protocols, i.e., frequency division multiple access (FDMA), TDMA, and NOMA. 
		These MOOPs are transformed into corresponding single-objective optimization problems (SOOPs) by invoking the $\epsilon$-constraint method. 
		For each multiple access method, we develop an effective particle swarm optimization (PSO)-based AO algorithm to jointly optimize PAs positions and resource allocation.
		\item
		Simulation results unveil that 
		1) compared to conventional fixed-position antenna systems, PASS can achieve significant rate-energy region gain; 
		2) by exploiting the time-switching characteristic for pinching beamforming, TDMA provides a higher performance gain than NOMA and FDMA in the multiple IUs/EUs scenario.
	\end{itemize}
	
	\subsection{Organization and Notation}
	
	The remainder of this paper is organized as follows.
	Section II studies the PASS-assisted SWIPT with single IU/EU, where a weighted sum optimization problem is formulated and a two-stage PA position optimization algorithm is proposed for the pinching beamforming design.
	Section III investigates the PASS-assisted SWIPT with multiple IUs/EUs, where MOOPs are formulated for three multiple access schemes.
	For each scheme, the corresponding joint PA position and resource allocation optimization algorithm is designed.
	Section IV provides simulation results and corresponding discussions.
	Finally, Section V concludes the paper.
	
	$\textit {Notations}$: Scalars, vectors, and matrices are denoted by italic letters, bold-face lower-case, and bold-face upper-case, respectively. 
	$\mathcal{CN}(\mu, \sigma^2)$ represents the distribution of a circularly symmetrical complex Gaussian random variable with a mean of $\mu$ and a variance of $\sigma^2$.
	Superscripts $(\cdot)^*$, $(\cdot)^T$, $(\cdot)^H$, and $(\cdot)^{-1}$ denote the conjugate, transpose, conjugate transpose, and inversion operators, respectively. 
	$|\cdot|$ and $\|\cdot\|$ denote the determinant and Euclidean norm of a vector, respectively.
	
	\section{PASS-Assisted SWIPT with Single IU/EU}
	
	In this section, we present the system model of the PASS-assisted SWIPT for the single IU/EU scenario, and formulate the PA position optimization problem.
	Then, we develop a two-stage pinching beamforming approach to explore the fundamental rate-energy region.
	
	\subsection{System Model}
	
	\begin{figure}[t]
		\centering
		\includegraphics[width=3.5in]{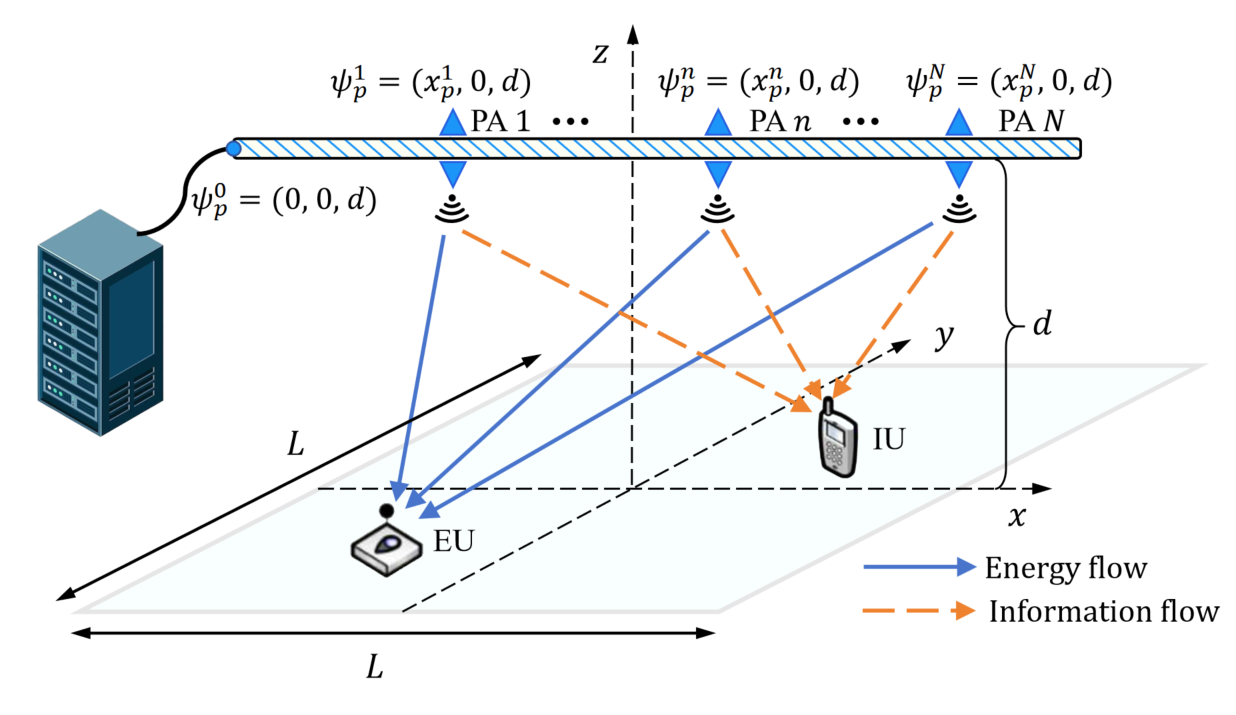}
		\caption{Illustration of the PASS-assisted SWIPT system with single IU and single EU.}
		\label{scenario1} 
	\end{figure}
	
	As illustrated in Fig.~\ref{scenario1}, a PASS-assisted SWIPT is considered, where the BS equipped with a single waveguide and $N$ PAs in a set $\mathcal{N} = \{1, ..., N\}$ serves a single-antenna IU and a single-antenna EU.
	In the three-dimensional Cartesian coordinate system, the dielectric waveguide is parallel to the $x$-axis with a height of $d$.
	The position of the $n$-th PA on the waveguide can be denoted as $\psi^n_\text{p} = \left[x^n_\text{p}, 0, d\right]^T$, 
	and the $x$-axis positions of all PAs can be represented by $\mathbf{x}_\text{p} = \left[x_\text{p}^1, ..., x_\text{p}^N\right]^T$.
	Furthermore, the feed point position of the waveguide is denoted as $\psi^0_\text{p} = \left[-\frac{L}{2}, 0, d\right]^T$.
	The maximum deployment range of PAs is $L$ and the service area of the PASS is a $L \times L$ square. 
	The positions of IU and EU are represented by $\psi_\mathrm{I} = \left[x_\mathrm{I}, y_\mathrm{I}, 0\right]^T$ and $\psi_\mathrm{E} = [x_\mathrm{E}, y_\mathrm{E}, 0]^T$, respectively.
	
	Since all PAs are located on the same waveguide, the signal transmitted by one PA is a phase-shifted version of the signal transmitted by the other PA, which means that one waveguide can only support one data stream.
	The propagation channel vector inside the waveguide can be expressed as
	\begin{equation}
		\begin{aligned}
			\mathbf{g}(\mathbf{x}_\text{p}) & = \left[e^{-j \frac{2\pi}{\lambda_g}\|\psi^{1}_\text{p} - \psi^{0}_\text{p}\|}, ..., e^{-j \frac{2\pi}{\lambda_g}\|\psi^{N}_\text{p} - \psi^{0}_\text{p}\|}\right]^T \\
			& = \left[e^{-j \frac{2\pi}{\lambda_g}\bar{x}^1_\text{p}}, ..., e^{-j \frac{2\pi}{\lambda_g}\bar{x}^N_\text{p}}\right]^T,
		\end{aligned}
	\end{equation}
	where $\bar{x}^n_\text{p} = x^n_\text{p} + \frac{L}{2}, \forall n \in \mathcal{N}$, $\lambda_g = \frac{\lambda}{n_\text{eff}}$ denotes the guided wavelength, and $n_\text{eff}$ is the effective refractive index of the dielectric waveguide~\cite{zhu2025pinching}.
	The free-space channel vector from PAs to the user $\text{u}$ is given by
	\begin{equation}
		\mathbf{h}_\text{u}(\mathbf{x}_\text{p})=\left[\frac{\eta^{\frac{1}{2}}e^{-j\frac{2\pi}{\lambda}\|\psi_\text{u}-{\psi}_\text{p}^{1}\|}}{\|\psi_\text{u}-{\psi_\text{p}^1}\|},\cdots,\frac{\eta^{\frac{1}{2}}e^{-j\frac{2\pi}{\lambda}\|\psi_\text{u}-{\psi}_\text{p}^N\|}}{\|\psi_\text{u}-{\psi}_\text{p}^N\|}\right]^T,
	\end{equation}
	where $\text{u}\in\{\text{I},\text{E}\}$ represents either IU or EU, $\eta = \frac{c^2}{16 \pi^2 f_c^2}$ is the path loss coefficient, $c$ is the light speed, $f_c$ is the signal frequency, and $\lambda$ is the signal wavelength.
	Moreover, $\|\psi_\text{u} - \psi^{n}_\text{p}\|$ is the distance between the $n$-th PA and the user $\text{u}$, which is given by
	\begin{equation}
		\begin{aligned}
			\|\psi_\text{u} - \psi^{n}_\text{p}\| & = 
			\sqrt{\left(x_\text{u} - x^n_\text{p} \right)^2 + y_\text{u}^2 + d^2}  \\
			& = \sqrt{\left(x_\text{u} - x^n_\text{p} \right)^2 + l_\text{u}^2},
		\end{aligned}
	\end{equation}
	where $l_\text{u} = \sqrt{y_\text{u}^2 + d^2}$ is a constant representing the distance between the $n$-th PA and user $\text{u}$ along the $y$-axis and $z$-axis.
	
	Let $s$ represent the normalized signal transmitted by PASS, with $\mathbb{E}\left[s s^*\right] = 1$.
	We assume that the total transmit power $P$ is equally distributed among $N$ PAs, leading to the transmit power of each PA being $\frac{P}{N}$.
	Therefore, the received signal of the user $\text{u}$ can be expressed as
	\begin{equation}
		y_\text{u} = \sqrt{\frac{P}{N}} \mathbf{h}^H_\text{u}\left(\mathbf{x}_\text{p}\right) \mathbf{g}\left(\mathbf{x}_\text{p}\right) s + n_\text{u},
	\end{equation}
	where $n_\text{u} \sim \mathcal{CN}(0, \sigma_\text{u}^2)$ is the additive white Gaussian noise (AWGN) at the user $\text{u}$, with $\sigma^2_\text{u}$ representing the noise power.
	As such, the received SNR for IU is given as
	\begin{equation}
		\label{single_snr}
		\begin{aligned}
			\mathrm{SNR}
			& = \frac{P}{N} \frac{\left|\mathbf{h}^H_\mathrm{I} \left(\mathbf{x}_\text{p}\right) \mathbf{g} \left(\mathbf{x}_\text{p}\right) \right|^2}{\sigma^2_\mathrm{I}} \\
			& = \frac{P \eta }{N \sigma^2_\mathrm{I}} \left|\sum^N_{n=1} \frac{e^{-j\left(\frac{2\pi}{\lambda}\|\psi_\mathrm{I}-{\psi}_\text{p}^{n}\|+ \frac{2\pi}{\lambda_g}\bar{x}^n_\text{p} \right)}}{\|\psi_\mathrm{I}-{\psi_\text{p}^{n}}\|}\right|^2.
		\end{aligned}
	\end{equation}
	Accordingly, the achievable rate of IU is expressed as $R = \log_2 \left(1+\mathrm{SNR}\right)$.
	Moreover, considering the linear energy harvesting model~\cite{ChenIRS2023}, the energy harvested by EU can be expressed as
	\begin{equation}
		\label{energy_single}
		\begin{aligned}
			E & = \zeta \frac{P}{N} \left|\mathbf{h}^H_\mathrm{E} \left(\mathbf{x}_\text{p}\right) \mathbf{g} \left(\mathbf{x}_\text{p}\right) \right|^2 \\
			& = \zeta \frac{P \eta }{N} \left|\sum^N_{n=1} \frac{e^{-j\left(\frac{2\pi}{\lambda}\|\psi_\mathrm{E}-{\psi}_\text{p}^{n}\|+ \frac{2\pi}{\lambda_g}\bar{x}^n_\text{p}\right)}}{\|\psi_\mathrm{E}-{\psi_\text{p}^{n}}\|}\right|^2,
		\end{aligned}
	\end{equation}
	where $\zeta \in (0, 1]$ is the energy conversion efficiency for EU.
	
	\subsection{Problem Formulation}
	In this paper, we aim to jointly maximize the data rate for IU and the harvested energy for EU by optimizing the pinching beamforming.
	Observing the SNR and energy harvesting expressions in \eqref{single_snr} and \eqref{energy_single}, we can find that the optimization problem can be equivalently transformed to the following one for maximizing the weighted effective channel gains to both IU and EU:
	\begin{subequations}
		\label{moopsingle}
		\begin{align}
			& \max_{\mathbf{x}_\text{p}}  \rho \left| \sum^N_{n=1} \frac{e^{j \phi^n_\mathrm{I}}}{\|\psi_\mathrm{I}-{\psi_\text{p}^{n}}\|} \right| + \left(1-\rho\right) \left| \sum^N_{n=1} \frac{e^{j \phi^n_\mathrm{E}}}{\|\psi_\mathrm{E}-{\psi_\text{p}^{n}}\|} \right|, \\
			\rm{s.t.} \ \
			& x^n_\text{p} - x^{n-1}_\text{p} \geq \Delta, \forall n, n-1 \in \mathcal{N}, \label{spacing1}\\
			& x^n_\text{p} \in \left[-L/2, L/2\right], \forall n \in \mathcal{N}, \label{range1}
		\end{align}
	\end{subequations}
	where $\rho \in \left[0, 1\right]$ is the weighted factor, $\phi^n_\mathrm{I} = \frac{2\pi}{\lambda}\|\psi_\mathrm{I}-{\psi}_\text{p}^{n}\|+ \frac{2\pi}{\lambda_g}\bar{x}^n_\text{p}$ and $\phi^n_\mathrm{E} = \frac{2\pi}{\lambda}\|\psi_\mathrm{E}-{\psi}_\text{p}^{n}\|+ \frac{2\pi}{\lambda_g}\bar{x}^n_\text{p}$ represent the phase shifts induced by the propagation within the waveguide and through free-space to the IU and EU, respectively.
	Constraint \eqref{spacing1} restricts the spacing among PAs to be no larger than $\Delta$ to avoid antenna coupling.
	Constraint \eqref{range1} gives the feasible deployment range for PAs on the waveguide.
	
	\subsection{Proposed Two-Stage Solution}
	\subsubsection{Problem Reformulation}
	Observing \eqref{moopsingle}, we can observe that the impact of pinching beamforming on SWIPT systems can be categorized into two aspects: large-scale path loss and signal phase shifts.  
	Based on this decomposition, we consider the following relaxed problem:
	\begin{subequations}
		\label{path0}
		\begin{align}
			& \max_{\mathbf{x}_\text{p}}  \rho  \sum^N_{n=1} \frac{1}{\|\psi_\mathrm{I}-{\psi_\text{p}^{n}}\|} + \left(1-\rho\right) \sum^N_{n=1} \frac{1}{\|\psi_\mathrm{E}-{\psi_\text{p}^{n}}\|} \label{path1}, \\
			\rm{s.t.} \ \
			& \phi^n_\mathrm{I} - \phi^{n-1}_\mathrm{I} = 2 k_1 \pi, \forall n, n-1 \in \mathcal{N}, k_1 \in \mathbb{Z}, \label{phiI1} \\
			& \phi^n_\mathrm{E} - \phi^{n-1}_\mathrm{E} = 2 k_2 \pi, \forall n, n-1 \in \mathcal{N}, k_2 \in \mathbb{Z}, \label{phiE1} \\
			& \eqref{spacing1}, \eqref{range1}.
		\end{align}
	\end{subequations}
	In \eqref{path0}, the objective function \eqref{path1} minimizes weighted path losses by maximizing the weighted sum of the reciprocals of the distances.
	Constraints \eqref{phiI1} and \eqref{phiE1} ensure phase alignment from all PAs to the IU and EU.
	Inspired by the fact that slight adjustments to PAs positions can result in periodic variations in signal phase shift, we propose a two-stage algorithm.   
	Specifically, coarse PAs positions are first determined by minimizing the large-scale path loss, which is followed by fine-tuning within a range of several wavelengths to achieve constructive signal combining.
	
	\subsubsection{Large-Scale Path Loss Minimization}
	
	With the objective of maximizing the large-scale path loss by neglecting phase alignment constraints \eqref{phiI1} and \eqref{phiE1}, we introduce auxiliary variables $\textbf{Y}_\mathrm{I} = \left[{Y}^{1}_\mathrm{I},..., {Y}^{N}_\mathrm{I}\right]^T$ and $\textbf{Y}_\mathrm{E} = \left[{Y}^{1}_\mathrm{E},..., {Y}^{N}_\mathrm{E}\right]^T$,  and obtain the following relaxed problem:
	\begin{subequations}
		 \label{path2}
		\begin{align}
			& \max_{\textbf{Y}_\mathrm{I}, \textbf{Y}_\mathrm{E}, \mathbf{x}_\text{p}^{\text{coarse}}} \sum^N_{n=1} \left(\rho \frac{1}{\sqrt{{Y}^{n}_\mathrm{I}}} 
			+ \left(1-\rho\right)  \frac{1}{\sqrt{{Y}^{n}_\mathrm{E}}}\right) \label{obj_path},\\
			\rm{s.t.} \ \
			& {Y}^{n}_\mathrm{I} \geq \left(x_\mathrm{I} - x^n_\text{p}\right)^2 + l_\mathrm{I}^2, \forall n \in \mathcal{N}, \label{YI}\\
			& {Y}^{n}_\mathrm{E} \geq \left(x_\mathrm{E} - x^n_\text{p}\right)^2 + l_\mathrm{E}^2, \forall n \in \mathcal{N}. \label{YE}\\
			& \eqref{spacing1}, \eqref{range1},
		\end{align}
	\end{subequations}
	where $\mathbf{x}_\text{p}^{\text{coarse}} = \left[x_\text{p}^{1, \text{coarse}}, ..., x_\text{p}^{N, \text{coarse}}\right]^T$ is the coarse PAs positions that minimize path loss.
	To tackle this non-convex problem, we adopt the SCA method.
	Specifically, the objective function \eqref{obj_path} can be linearized via the first-order Taylor expansion at the given local points $Y_\mathrm{I}^{n,\left(k\right)}$ and $Y_\mathrm{E}^{n,(k)}$ as follows:
	\begin{subequations}
		\begin{align}
			& \frac{1}{\sqrt{Y^{n}_\mathrm{I}}} \geq \frac{1}{\sqrt{Y_\mathrm{I}^{n,\left(k\right)}}} - \frac{\left( Y^{n}_\mathrm{I} - Y_\mathrm{I}^{n,(k)} \right)}{2 \left( Y_\mathrm{I}^{n,\left(k\right)} \right)^{3/2}} \triangleq \text{A}^{n}_\mathrm{I},\\
			& \frac{1}{\sqrt{Y^{n}_\mathrm{E}}} \geq \frac{1}{\sqrt{Y_\mathrm{E}^{n,\left(k\right)}}} - \frac{\left( Y^{n}_\mathrm{E} - Y_\mathrm{E}^{n,(k)} \right)}{2 \left( Y_\mathrm{E}^{n,\left(k\right)} \right)^{3/2}} \triangleq \text{A}^{n}_\mathrm{E}.
		\end{align}
	\end{subequations}
	By replacing the non-concave objective function in the problem \eqref{path2} with its concave lower bound, the optimization problem can be relaxed as
	\begin{subequations}
		\label{path3}
		\begin{align}
			& \max_{\textbf{Y}_\mathrm{I}, \textbf{Y}_\mathrm{E}, \mathbf{x}_\text{p}^{\text{coarse}}}  \sum_{n=1}^{N} \left[ \rho  \text{A}^{n}_\mathrm{I} + \left(1 - \rho\right) \text{A}^{n}_\mathrm{E} \right],\\
			\rm{s.t.} \ \
			& \eqref{spacing1}, \eqref{range1}, \eqref{YI}, \eqref{YE}.
		\end{align}
	\end{subequations}
	Problem \eqref{path3} is a convex optimization problem, which can be efficiently solved by standard convex solvers such as CVX~\cite{grant2014cvx}.
	
	\subsubsection{{Phase Alignment}}
	
	After obtaining the coarse PAs positions that minimize path loss, we use the fine-tuning method to achieve constructive combinations for signals received at both IU and EU.
	To ensure the accuracy of the phase approximation and to avoid significant deviation from the optimized path loss, only slight adjustments are made to the positions of PAs. 
	Thus, the phase $\phi^n_\text{u}, \text{u}\in\{\text{I},\text{E}\}$ can be approximated around its neighboring PA phase $\phi^{n-1}_\text{u}$ by first-order Taylor expansion, which is given by
	\begin{equation}
		\label{taylorphi1}
		\begin{aligned}
			\phi^n_\text{u} &\approx \frac{2\pi}{\lambda} \frac{x^{n-1}_\text{p} - x_\text{u}}{\sqrt{\left(x^{n-1}_\text{p} - x_\text{u}\right)^2 + l_\text{u}^2}} \left(x^n_\text{p} -x^{n-1}_\text{p}\right) \\
			& \quad + \frac{2\pi}{\lambda} \sqrt{\left(x^{n-1}_\text{p} - x_\text{u}\right)^2 + l_\text{u}^2} + \frac{2\pi}{\lambda_g} \bar{x}^n_\text{p} \\
			& = \frac{2\pi}{\lambda} \frac{x^{n-1}_\text{p} - x_\text{u}}{\sqrt{\left(x^{n-1}_\text{p} - x_\text{u}\right)^2 + l_\text{u}^2}} \left(x^n_\text{p} -x^{n-1}_\text{p}\right) \\
			& \quad + \phi^{n-1}_\text{u} + \frac{2\pi}{\lambda_g} (x^n_\text{p} - x^{n-1}_\text{p}).
		\end{aligned}
	\end{equation}
	Observing that \eqref{taylorphi1} indicates the relationship between $\left(\phi^n_\text{u} - \phi^{n-1}_\text{u}\right)$ and $\left(x^n_\text{p} -x^{n-1}_\text{p}\right)$.
	Substituting \eqref{taylorphi1} into constraints \eqref{phiI1} and \eqref{phiE1} yields
	\begin{subequations}
		\begin{align}
			& \Delta_x(\frac{1}{\lambda} \frac{x^{n-1}_\text{p} - x_\mathrm{I}}{\sqrt{\left(x^{n-1}_\text{p} - x_\mathrm{I}\right)^2 + l_\mathrm{I}^2}} + \frac{1}{\lambda_g}) = k_1, \\
			& \Delta_x(\frac{1}{\lambda} \frac{x^{n-1}_\text{p} - x_\mathrm{E}}{\sqrt{\left(x^{n-1}_\text{p} - x_\mathrm{E}\right)^2 + l_\mathrm{E}^2}} + \frac{1}{\lambda_g}) = k_2,
		\end{align}
	\end{subequations}
	where $\Delta_x = x^n_\text{p} -x^{n-1}_\text{p}$, which can be calculated as
	\begin{equation}
		\label{deltax1}
		\Delta_x = \frac{k_1 + k_2}{\frac{1}{\lambda}\left(\frac{x^{n-1}_\text{p} - x_\mathrm{I}}{\sqrt{\left(x^{n-1}_\text{p} - x_\mathrm{I}\right)^2 + l_\mathrm{I}^2}} + \frac{x^{n-1}_\text{p} - x_\mathrm{E}}{\sqrt{\left(x^{n-1}_\text{p} - x_\mathrm{E}\right)^2 + l_\mathrm{E}^2}}\right) + \frac{2}{\lambda_g}}.
	\end{equation}
	Therefore, when $x^{n-1}_\text{p}$ is given, the position of the $n$-th PA can be obtained by $x^n_\text{p} = x^{n-1}_\text{p} + \Delta_x$.
	Moreover, when $x^n_\text{p}$ is given, we can get the position of the $\left(n-1\right)$-th PA by $x^{n-1}_\text{p} = x^n_\text{p} - \Delta_{x'}$, where $\Delta_{x'}$ can be expressed as
	\begin{equation}
		\label{deltax2}
		\Delta_{x'} = \frac{k_1 + k_2}{\frac{1}{\lambda}\left(\frac{x^{n}_\text{p} - x_\mathrm{I}}{\sqrt{\left(x^{n}_\text{p} - x_\mathrm{I}\right)^2 + l_\mathrm{I}^2}} + \frac{x^{n}_\text{p} - x_\mathrm{E}}{\sqrt{\left(x^{n}_\text{p} - x_\mathrm{E}\right)^2 + l_\mathrm{E}^2}}\right) + \frac{2}{\lambda_g}}.
	\end{equation}
    Meanwhile, the step sizes $\Delta_{x}$ and $\Delta_{x'}$ should satisfy $\Delta_{x} \geq \Delta$ and $\Delta_{x'} \geq \Delta$ to avoid the coupling effect.
	With $\Delta_x$ and $\Delta_{x'}$, we need to find the reference PA for the fine-tuning. Specifically, we can choose the PA position with the minimum weighted path loss, i.e., $x_\text{p}^{n^*}$, as the reference one, that is,
	\begin{equation}
		\label{obj}
		x_\text{p}^{n^*} = \operatorname{argmax}_{\mathbf{x}^{\text{coarse}}_\text{p}} \rho \frac{1}{\|\psi_\mathrm{I}-{\psi_\text{p}^{n}}\|} + \left(1 - \rho\right) \frac{1}{\|\psi_\mathrm{E}-{\psi_\text{p}^{n}}\|},
	\end{equation}
	where $n^*$ is the index of the PA among the set of PAs.
	The positions of PAs to the left ($n < n^*$) are determined by $x^n_\text{p} = x^{n+1}_\text{p} - \Delta_{x'}$, while those to the right ($n > n^*$) are adjusted using $x^n_\text{p} = x^{n-1}_\text{p} + \Delta_{x}$.
	Note that, to preserve the path loss determined in the first stage, we select the fine-tuned PA position $x^n_\text{p}$ closest to the original coarse PA position $x^{n, \text{coarse}}_\text{p}$, while satisfying the antenna spacing constraint.
	
	The overall two-stage algorithm for solving problem \eqref{path0} is shown in \textbf{Algorithm~\ref{lowcomplex}}.
	Using the interior-point method, the computational complexity of the first-stage SCA algorithm is $\mathcal{O}(I_{\text{iter}} N^{3.5})$, where $I_{\text{iter}}$ is the number of SCA iterations.
	For the second stage, the computational complexity of the fine-tuning algorithm is $\mathcal{O}(N)$.
	Therefore, the overall complexity of the proposed algorithm is $\mathcal{O}(I_{\text{iter}} N^{3.5})$.
	
	\begin{algorithm}[t]
			\caption{Proposed Two-Stage Algorithm to Solve Problem \eqref{path0}}
			\label{lowcomplex}
			\begin{algorithmic}[1]
					\STATE Initialize feasible $\{\mathbf{Y}_{\mathrm{I}}^{(0)}, \mathbf{Y}_{\mathrm{E}}^{(0)}, \mathbf{x}_\text{p}^{(0)}\}$ and convergence criterion $\varepsilon_1$. Set $k = 1$.
					\REPEAT
					\STATE Update $\{\mathbf{Y}_{\mathrm{I}}^{(k)}, \mathbf{Y}_{\mathrm{E}}^{(k)}, \mathbf{x}_\text{p}^{(k)}\}$ by solving \eqref{path3}.
					\STATE $k = k + 1$.
					\UNTIL{the fractional increment of the objective function is lower than the threshold $\varepsilon_1$.}
					\STATE Find the optimal position $x^{n^*}_\text{p}$ of PA with \eqref{obj}.
					\FOR{$n=1:N$} 
					\IF{$n<n^*$}
					\STATE Update $x^{n^*-n}_\text{p}$ by $x^{n^*-n+1}_\text{p} - \Delta_{x'}$ using \eqref{deltax2}.
					\ELSIF{$n>n^*$}
					\STATE Update $x^{n}_\text{p}$ by $x^{n-1}_\text{p} + \Delta_{x}$ using \eqref{deltax1}.
					\ENDIF
					\ENDFOR
					\STATE Output PAs positions $\mathbf{x}_\text{p}$.
				\end{algorithmic}
		\end{algorithm}

	\section{PASS-Assisted SWIPT with Multiple IUs/EUs}
	
	In this section, we present the system models for the multiple IUs/EUs scenario under three multiple access schemes. 
	To ensure user fairness, we formulate MOOPs that simultaneously maximize the minimum data rate for IU and the minimum harvested energy for EU. 
	To solve these problems, we employ the $\epsilon$-constraint method to convert them into SOOPs, and then propose the PSO-based AO algorithms to jointly optimize the pinching beamforming and resource allocation.
	
	\subsection{System Model and Problem Formulation}
	
	\begin{figure}[t]
		\centering
		\includegraphics[width=3.5in]{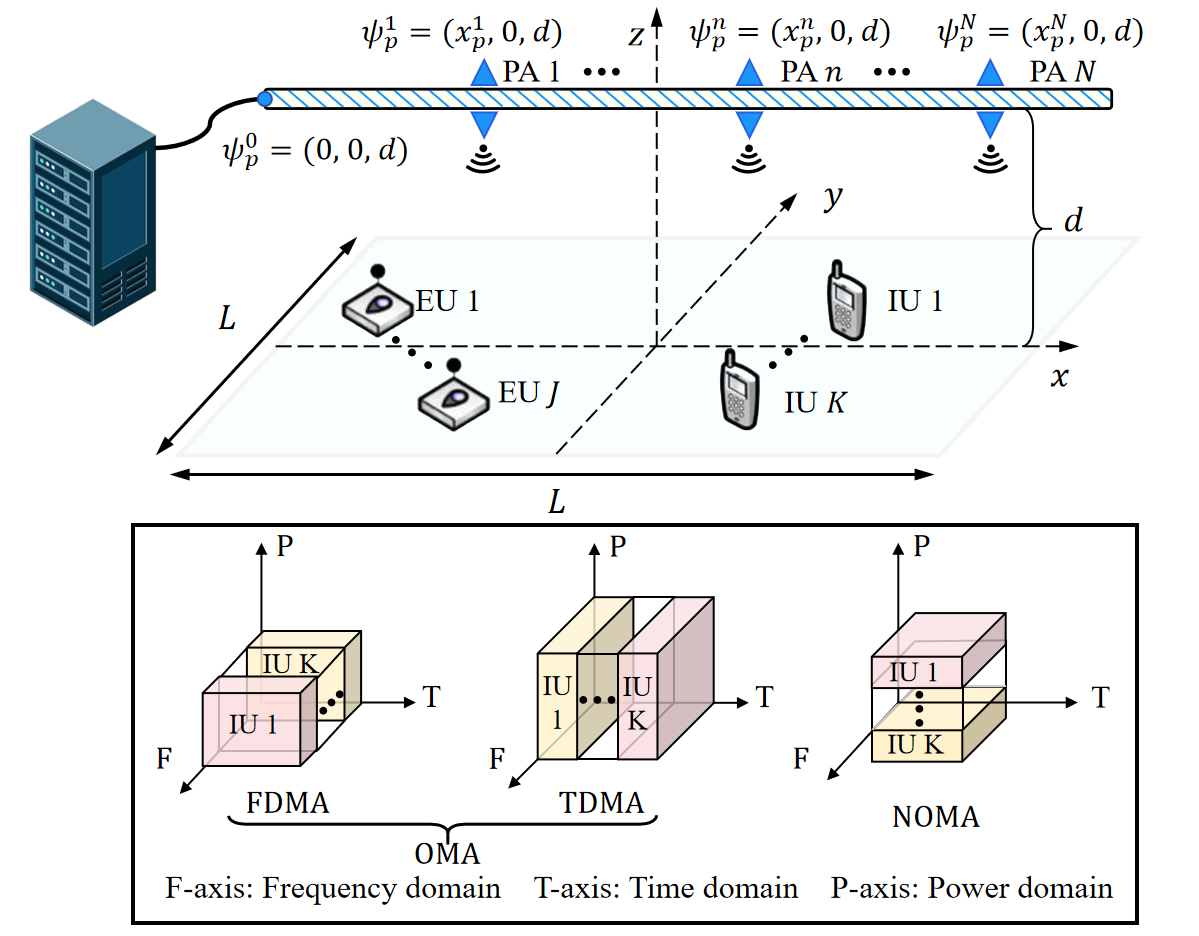}
		\caption{Illustration the PASS-assisted SWIPT system with multiple IUs and EUs.}
		\label{scenario2}
	\end{figure}
	The considered scenario is shown in Fig.~\ref{scenario2}, where $N$ PAs on a single waveguide are deployed to serve two sets of single-anenna users, i.e., IUs and EUs.
	The sets of IUs and EUs are denoted by $\mathcal{K} = \{\mathrm{I}_1, ..., \mathrm{I}_K\}$ and $\mathcal{J} = \{\mathrm{E}_1, ..., \mathrm{E}_J\}$, respectively.
	The signals received by the $k$-th IU and $j$-th EU are respectively given as
	\begin{subequations}
	\begin{align}
		& y_{\mathrm{I},k} = \sqrt{\frac{P}{N}}\mathbf{h}^H_{\mathrm{I},k} \left(\mathbf{x}_\text{p}\right) \mathbf{g} \left(\mathbf{x}_\text{p}\right) s + n_{\mathrm{I},k}, \\
		& y_{\mathrm{E},j} = \sqrt{\frac{P}{N}}\mathbf{h}^H_{\mathrm{E},j} \left(\mathbf{x}_\text{p}\right) \mathbf{g} \left(\mathbf{x}_\text{p}\right) s + n_{\mathrm{E},j},
		\end{align}
	\end{subequations}
	where $\mathbf{h}_{\mathrm{I},k}$ and $\mathbf{h}_{\mathrm{E},j}$ denote the channels from PAs to the $k$-th IU and the $j$-th EU, respectively, $n_{\mathrm{I},k} \sim \mathcal{CN}(0, \sigma_{\mathrm{I}, k}^2)$ and $n_{\mathrm{E},j} \sim \mathcal{CN}(0, \sigma_{\mathrm{E}, j}^2)$ denote the AWGNs at the $k$-th IU and the $j$-th EU, with $\sigma_{\mathrm{I}, k}^2$ and $\sigma_{\mathrm{E}, j}^2$ denoting the noise powers, respectively.
	
	Since a single waveguide can only transmit one data stream, we consider three different multiple access schemes, i.e., FDMA, TDMA, and NOMA for serving multiple IUs and EUs.
	
	\subsubsection{FDMA-Assisted PASS}
	
	For FDMA, the frequency band is divided into multiple orthogonal frequency slots to serve multiple IUs.
	The achievable rate for the $k$-th IU is then given by
	\begin{equation}
		R_k = w_k \log_2 \left(1+\frac{p_k \left|\mathbf{h}^H_{\mathrm{I},k} \left(\mathbf{x}_\text{p}\right) \mathbf{g} \left(\mathbf{x}_\text{p}\right) \right|^2}{N w_k \sigma^2_{\mathrm{I},k}}\right),
	\end{equation}
	where $w_k$ and $p_k$ denote the frequency and power allocation factor of the $k$-th IU, respectively.
	
	Note that the decoding order and power allocation in FDMA have no effect on the EUs' energy harvested.
	Thus, the energy harvested by the $j$-th EU can be expressed as
	\begin{equation}
		\label{e_j_noma}
		E_j = \zeta \frac{P}{N} \left|\mathbf{h}^H_{\mathrm{E},j} \left(\mathbf{x}_\text{p}\right) \mathbf{g} \left(\mathbf{x}_\text{p}\right)\right|^2.
	\end{equation}
	Observing \eqref{e_j_noma}, we can find that the harvested energy only depends on the channel gain that is determined by the PAs positions.
	
	Our aim is to simultaneously maximize the minimum data rate for IUs and the minimum harvested energy for EUs to guarantee user fairness, for which the MOOP is formulated as
	\begin{subequations}
		\label{moop_fdma}
		\begin{align}
			(\text{FDMA}): & \max_{\mathbf{x}_\text{p}, \mathbf{w}, \mathbf{p}}  \min_{k \in \mathcal{K}}  R_{k},\\
			& \max_{\mathbf{x}_\text{p}}  \min_{j \in \mathcal{J}}  E_j,\\
			\rm{s.t.} \ \
			& \sum_{k=1}^{K} w_k \leq 1, \forall k \in \mathcal{K}, \label{w1}\\
			& 0 \leq w_k \leq 1, \forall k \in \mathcal{K}, \label{w2}\\
			& \sum_{k=1}^{K} p_k \leq P, \forall k \in \mathcal{K}, \label{pk_fdma}\\
			& \eqref{spacing1}, \eqref{range1},
		\end{align}
	\end{subequations}
	where $\mathbf{w} = \left[w_1, ..., w_K\right]^T$ and $\mathbf{p} = \left[p_1, ..., p_K\right]^T$ denote the frequency and power allocation vectors, respectively.
	\subsubsection{TDMA-Assisted PASS}
	For TDMA, PAs serve different IUs through different orthogonal time slots.
	Given the flexible positioning of the PAs, we develop a time-switching operation mode for the PASS to fully exploit its potential. 
	Specifically, the PAs can be redeployed within each time slot to provide service for one IU and all EUs.
	Thus, the achievable rate for the $k$-th IU in its allocated $k$-th time slot is given by
	\begin{equation}
		R_k = \tau_k \log_2 \left(1+\frac{P}{N \sigma^2_{\mathrm{I},k}} \left|\mathbf{h}^H_{\mathrm{I},k} \left(\mathbf{x}_{\text{p}, k}\right) \mathbf{g} \left(\mathbf{x}_{\text{p}, k}\right)\right|^2\right),
	\end{equation}
	where $\tau_k$ is the time allocation factor for the $k$-th IU, and $\mathbf{x}_{\text{p}, k} = \left[x^1_{\text{p},k}, ..., x^N_{\text{p},k}\right]^T$ denotes the positions of PAs in the $k$-th time slot.
	For the $j$-th EU, its harvested energy in the $k$-th time slot is given by
	\begin{equation}
		E^{k}_{j} = \tau_k \zeta \frac{P}{N} \left|\mathbf{h}^H_{\mathrm{E},j}\left(\mathbf{x}_{\text{p}, k}\right) \mathbf{g}\left(\mathbf{x}_{\text{p}, k}\right)\right|^2.
	\end{equation}
	Then, the total harvested energy by the $j$-th EU can be expressed as
	\begin{equation}
		E_j = \zeta \frac{P}{N} \sum_{k=1}^{K} \tau_k \left|\mathbf{h}^H_{\mathrm{E},j}\left(\mathbf{x}_{\text{p}, k}\right) \mathbf{g}\left(\mathbf{x}_{\text{p}, k}\right)\right|^2,
	\end{equation}
	where the harvested energy depends on both pinching beamforming and time allocation.
	Thus, the corresponding MOOP for TDMA can be written as
	\begin{subequations}
		\label{moop_tdma}
		\begin{align}
			(\text{TDMA}):  & \max_{\mathbf{X}_\text{p}, \boldsymbol{\tau}}  \min_{k \in \mathcal{K}}  R_{k},\\
			& \max_{\mathbf{X}_\text{p}, \boldsymbol{\tau}}  \min_{j \in \mathcal{J}}  E_j,\\
			\rm{s.t.} \ \
			& x^n_{\text{p},k} - x^{n-1}_{\text{p},k} \geq \Delta, \forall n, n-1 \in \mathcal{N}, \forall k \in \mathcal{K}, \label{spacing2}\\
			& x^n_{\text{p},k} \in [-L/2, L/2], \forall n \in \mathcal{N}, \forall k \in \mathcal{K}, \label{range2}\\
			& \sum_{k=1}^{K} \tau_k \leq 1, \forall k \in \mathcal{K}, \label{tau1}\\
			& 0 \leq \tau_k \leq 1, \forall k \in \mathcal{K}, \label{tau2}
		\end{align}
	\end{subequations}
	where $\mathbf{X}_\text{p} = \left[\mathbf{x}_{\text{p}. 1}, ... , \mathbf{x}_{\text{p}. K}\right] \in \mathbb{R}^{N \times K}$ and $\boldsymbol{\tau} = \left[\tau_1, ..., \tau_K\right]^T$ denote the position matrix of PAs and the time allocation vector, respectively.
	
	\subsubsection{NOMA-Assisted PASS}
	
	For NOMA, all IUs are served over the same time/frequency resource block, with user multiplexing performed in the power domain.
	The superimposed signal transmitted is $s = \sum_{k=1}^{K} \sqrt{\alpha_k} s_k$, where $\alpha_k$ and $s_k$ are the power allocation coefficient and the expected signal of the $k$-th IU, respectively. 
	According to the SIC principle, the IU with strong channel gain first decodes and removes the signals intended for the IUs with weak channel gains, and then decodes its own signal.
	We define $\boldsymbol{\pi} = \{\pi(1), ..., \pi(K)\}$ as the decoding orders of all IUs.
	For any $k$ and $k'$, if $\pi(k) > \pi(k')$, it implies that the $k$-th IU first decodes the signal of the $k'$-th IU before decoding its own signal.
	Consequently, their combined channel gains must satisfy $\left|\mathbf{h}^H_{\mathrm{I},k} \left(\mathbf{x}_\text{p}\right) \mathbf{g} \left(\mathbf{x}_\text{p}\right)\right|^2 \geq \left|\mathbf{h}^H_{\mathrm{I},k'} \left(\mathbf{x}_\text{p}\right) \mathbf{g} \left(\mathbf{x}_\text{p}\right)\right|^2$.
	
	Thus, the achievable rate for the $k$-th IU is denoted as
	\begin{equation}
		R_{k} = \log_2 \left(1+\frac{\alpha_k \frac{P}{N} \left|\mathbf{h}^H_{\mathrm{I},k} \left(\mathbf{x}_\text{p}\right) \mathbf{g} \left(\mathbf{x}_\text{p}\right)\right|^2}{\sum_{\pi(i) > \pi(k)} \alpha_i \frac{P}{N} \left|\mathbf{h}^H_{\mathrm{I},k} \left(\mathbf{x}_\text{p}\right) \mathbf{g} \left(\mathbf{x}_\text{p}\right)\right|^2 +\sigma^2_{\mathrm{I},k}} \right),
	\end{equation}
	where $\sigma^2_{\mathrm{I},k}$ is the noise power for the $k$-th IU.
	Furthermore, for a given decoding order, the power allocation coefficients should satisfy the following condition:
	\begin{equation}
		0 \leq \alpha_k \leq \alpha_{k'}, \text{if } \pi(k) > \pi(k'), \forall k, k' \in \mathcal{K}.
	\end{equation}
	
	Since the power and frequency allocation of NOMA have no effect on the energy harvesting of EUs, the energy harvested by the $j$-th EU is the same as that given in \eqref{e_j_noma}.
	Accordingly, the MOOP for NOMA can be formulated as
	\begin{subequations}
		\label{moop_noma}
		\begin{align}
			(\text{NOMA}): & \max_{\mathbf{x}_\text{p}, \mathbf{a}}  \min_{k \in \mathcal{K}}  R_{k}, \\
			 & \max_{\mathbf{x}_\text{p}}  \min_{j \in \mathcal{J}}  E_j, \\
			\rm{s.t.} \ \
			& \notag\left|\mathbf{h}^H_{\mathrm{I},k} \left(\mathbf{x}_\text{p}\right) \mathbf{g} \left(\mathbf{x}_\text{p}\right)\right|^2 \geq \left|\mathbf{h}^H_{\mathrm{I},k'} \left(\mathbf{x}_\text{p}\right) \mathbf{g} \left(\mathbf{x}_\text{p}\right)\right|^2, \\
			& \text{if } \pi(k) > \pi(k'), \forall k, k' \in \mathcal{K}, \label{gain}\\
			& 0 \leq \alpha_k \leq \alpha_{k'},  \text{if } \pi(k) > \pi(k'), \label{alpha1}\\
			& \sum_{k=1}^{K} \alpha_k \leq 1, \forall k \in \mathcal{K}, \label{alpha2}\\
			& \eqref{spacing1}, \eqref{range1},
		\end{align}
	\end{subequations}
	where $\mathbf{a} = [\alpha_1, ..., \alpha_K]^T$ is the power allocation coefficient vector.

	Due to the conflicts among multiple objectives and the strong coupling among involved variables, problems \eqref{moop_fdma}, \eqref{moop_tdma}, and \eqref{moop_noma} are challenging to solve.  
	In the following, for the MOOP under each multiple access scheme, we first invoke the $\epsilon$-constrant method to reformulate the problem into a more tractable one, and then utilize the PSO and convex optimization algorithms for pinching beamforming and resource allocation, respectively.
	
	\subsection{Proposed Solutions}
	
	\subsubsection{FDMA-Assisted PASS}
	
	Inspired by the capacity of $\epsilon$-constraint method to effectively explore the Pareto frontier of complex conflicting objectives, we adopt it to transform the MOOP \eqref{moop_fdma} into an SOOP.
	Specifically, the objective of maximizing the minimum energy harvested by EUs is converted into a constraint related to $\epsilon$, yielding the following SOOP:
	\begin{subequations}
		\label{soop_fdma}
		\begin{align}
			& \max_{\mathbf{x}_\text{p}, \mathbf{w}, \mathbf{p}}  \min_{k \in \mathcal{K}}  R_{k}, \\
			\rm{s.t.} \ \
			& E_j \geq \epsilon, \forall j \in \mathcal{J}, \label{e_j_fdma} \\
			& \eqref{spacing1}, \eqref{range1}, \eqref{w1}, \eqref{w2}, \eqref{pk_fdma},
		\end{align}
	\end{subequations}
	where constraint \eqref{e_j_fdma} is the minimum harvested energy requirement for EUs.
	However, due to the mutual coupling between the PAs positions and resource allocation coefficients, this problem remains difficult to solve.
	Therefore, we decompose the original problem into two subproblems, and utilize the AO algorithm to iteratively optimize the positions of PAs $\mathbf{x}_p$ and resource allocation vectors $\mathbf{w}, \mathbf{p}$.
	
	\textbf{Optimizing $\{\mathbf{x}_\text{p}\}$}:
	With the frequency and power allocation vectors $\mathbf{w}$ and $\mathbf{p}$ fixed, problem~\eqref{soop_fdma} can be reformulated as
	\begin{subequations}
		\label{soop_fdma_x}
		\begin{align}
			& \max_{\mathbf{x}_\text{p}}  \min_{k \in \mathcal{K}}  R_{k}, \\
			\rm{s.t.} \ \
			& \eqref{spacing1}, \eqref{range1}, \eqref{e_j_fdma},
		\end{align}
	\end{subequations}
	Given the high non-convexity of the problem with respect to the PAs positions and its large solution space, obtaining the optimal solutions via conventional optimization methods is challenging.
	To tackle this problem, we employ the PSO algorithm~\cite{Clercparticle2002} to optimize the PAs positions.
	To begin with, we first initialize a particle swarm $\mathcal{S} = \{1, ... , S\}$, where each particle is associated with the positions of all PAs.
	The initial position of the $s$-th particle can be expressed as
	\begin{equation}
		\mathbf{x}^{(0)}_s = \left[x^{(0)}_{s, 1}, x^{(0)}_{s, 2}, ..., x^{(0)}_{s, N}\right]^T,
	\end{equation}
	where $x^{(0)}_{s, n}$ represents the $x$-coordinate of the $n$-th PA. 
	To guarantee the constraints \eqref{spacing1} and \eqref{range1}, PAs positions are randomly distributed within $\left[-L/2, L/2\right]$, while ensuring a minimum PA spacing of $\lambda/2$.
	Accordingly, the initial velocity of the $s$-th particle is defined as
	\begin{equation}
		\mathbf{v}^{(0)}_s = \left[v^{(0)}_{s, 1}, v^{(0)}_{s, 2}, ..., v^{(0)}_{s, N}\right]^T.
	\end{equation}
	
	\begin{algorithm}[t]
		\caption{Proposed PSO Algorithm to Solve Problem \eqref{soop_fdma_x}}
		\label{solve_x_p}
		\begin{algorithmic}[1]
			\STATE Initialize the particle swarm with $\{\mathbf{x}_s^{(0)}, \mathbf{v}_s^{(0)}, \forall s \in \mathcal{S}\}$. Set the convergence criterion $\varepsilon_2$.
			\STATE Set the personal best position $\{\mathbf{x}_{s,\text{p}} = \mathbf{x}_s^{(0)}, \forall s \in \mathcal{S}\}$ for each particle, and the global best postion $\mathbf{x}_\text{g} = \operatorname{argmax}_{\mathbf{x}_s^{(0)}}\{\mathcal{F}(\mathbf{x}_{s,\text{p}}), \forall s \in \mathcal{S}\}$.
			\REPEAT
			\STATE Set the iteration index $l = 1$.
			\REPEAT
			\STATE Update $\omega$ with \eqref{omega}, and set the particle index $s = 1$.
			\REPEAT
			\STATE Update the velocity and position of the $s$-th particle according to \eqref{update_v} and \eqref{update_x}, respectively.
			\STATE Evaluate the fitness function value $\mathcal{F}(\mathbf{x}_s^{(l)})$ for the $s$-th particle according to \eqref{fit}.
			\IF{$\mathcal{F}(\mathbf{x}_s^{(l)}) > \mathcal{F}(\mathbf{x}_{i,\text{p}})$}
			\STATE Update $\mathbf{x}_{s,\text{p}} = \mathbf{x}_s^{(l)}$.
			\ENDIF
			\IF{$\mathcal{F}(\mathbf{x}_s^{(l)}) > \mathcal{F}(\mathbf{x}_\text{g})$}
			\STATE Update $\mathbf{x}_\text{g} = \mathbf{x}_s^{(l)}$.
			\ENDIF
			\STATE $s = s + 1$.
			\UNTIL{$s > \mathcal{S}$}.
			\STATE $l = l + 1$.
			\UNTIL{$l > L_{\max}$}.
			\UNTIL{the fractional increment of the objective function is below $\varepsilon_2$}.
		\end{algorithmic}
	\end{algorithm}
	
	Then, by introducing the penalty factors, we transform problem \eqref{soop_fdma_x} into an unconstrained problem as the fitness function, which is designed as
	\begin{equation}
		\label{fit}
		\mathcal{F}\left(\mathbf{x}^{(l)}_s\right) = \xi - \rho_1 \mathcal{P}_1 \left(\mathbf{x}^{(l)}_s\right) - \rho_2 \mathcal{P}_2 \left(\mathbf{x}^{(l)}_s\right)
		- \rho_3 \mathcal{P}_3 \left(\mathbf{x}^{(l)}_s\right),
	\end{equation}
	where $\xi = \min(R_k), \forall k \in \mathcal{K}$, $l$ denotes the iterative index,  and $\rho_1$, $\rho_2$, and $\rho_3$ are the penalty factors.
	To guarantee the constraints of the optimization problem, the penalty factors are generally set as sufficiently large values. 
	$\mathcal{P}_1 (\mathbf{x}^{(l)}_s)$, $\mathcal{P}_2 (\mathbf{x}^{(l)}_s)$, and $\mathcal{P}_3 (\mathbf{x}^{(l)}_s)$ represent the penalty functions of the current PAs positions that prevent from violating the constraints, which are given by~\cite{Clercparticle2002}
	\begin{equation}
		\label{P1X}
		\mathcal{P}_1(\mathbf{x}^{(l)}_s) = \sum_{n=2}^{N} \mathbb{I}\left(\|x^{(l)}_{s,n} - x^{(l)}_{s,n-1}\| \geq \Delta\right),
	\end{equation}
	\begin{equation}
		\label{P2X}
		\mathcal{P}_2(\mathbf{x}^{(l)}_s) = \sum_{n=1}^{N} \mathbb{I}\left(x^{(l)}_{s,n} \in \left(-\infty, -\frac{L}{2}\right) \bigcup \left(\frac{L}{2}, \infty\right)\right),
	\end{equation}
	\begin{equation}
		\label{P3X}
		\mathcal{P}_3(\mathbf{x}^{(l)}_s) = 
		\begin{cases}
			\epsilon - \min(E_{j}), &\text{if} \min(E_{j}) < \epsilon, \\
			0, &\text{otherwise},
		\end{cases}
	\end{equation}
	where $\mathbb{I}(\cdot)$ represents an indicator function that equals one if the condition within the parentheses holds and otherwise equals zero.
	Moreover, equation \eqref{P1X} ensures that the minimum separation between PAs exceeds the coupling distance, equation \eqref{P2X} enables the PAs to be distributed within the effective waveguide range, and equation \eqref{P3X} guarantees the harvested energy threshold requirement.
	During the iterative process, each particle dynamically updates its personal best position and global best position based on the fitness function~\cite{CoelloHandling2004}.
	Let $\mathbf{x}_{s,\text{p}}$ and $\mathbf{x}_\text{g}$ denote the personal best position of the $s$-th particle and the global best position of entire particles.
	In each iteration, the update equations for position and velocity are given by
	\begin{equation}
		\label{update_v}
		\mathbf{v}^{(l+1)}_{s} = \omega \mathbf{v}^{(l)}_{s} + c_1 r_1 \left(\mathbf{x}_{s, \text{p}} - \mathbf{x}^{(l)}_s\right) + c_2 r_2\left(\mathbf{x}_\text{g} - \mathbf{x}^{(l)}_s\right),
	\end{equation}
	\begin{equation}
		\label{update_x}
		\mathbf{x}^{(l+1)}_{s} = \mathbf{x}^{(l)}_s + \mathbf{v}^{(l+1)}_{s},
	\end{equation}
	where $\omega$ denotes the inertia representing the degree of trust in the previous velocity direction, $c_1$ and $c_2$ are acceleration constants used to adjust the maximum learning step size, and $r_1$ and $r_2$ are two random numbers in the range $[0, 1]$ for improving the randomness of the search.
	Moreover, $\omega$ is updated with the following criteria:
	\begin{equation}
		\label{omega}
		\omega = \omega_{\max} - \frac{\left(\omega_{\max} - \omega_{\min}\right) \cdot l}{L_{\max}},
	\end{equation}
	where $\omega_{\max}$ and $\omega_{\min}$ are the maximum and minimum inertial weights, and $L_{\max}$ is the maximum iteration.
	The details of the proposed PSO algorithm are shown in \textbf{Algorithm~\ref{solve_x_p}}.
	
	\textbf{Optimizing $\{\mathbf{w}, \mathbf{p}\}$}:
	Given the specified PAs positions $\mathbf{x}_\text{p}$, we then optimize the frequency allocation vector $\mathbf{w}$ and power allocation vector $\mathbf{p}$ by solving the following subproblem:
	\begin{subequations}
		\begin{align}
			& \max_{\mathbf{w}, \mathbf{p}} \min_{k \in \mathcal{K}}  R_{k}, \\
			\rm{s.t.} \ \
			& \eqref{w1}, \eqref{w2}, \eqref{pk_fdma}.
		\end{align}
	\end{subequations}
	By introducing the auxiliary variable $\xi$ to denote the minimum achievable rate, the subproblem is given by
	\begin{subequations}
		\begin{align}
			& \max_{\mathbf{w}, \mathbf{p}, \xi}  \xi,  \\
			\rm{s.t.} \ \
			& \xi - R_k \leq 0, \forall k \in \mathcal{K}, \\
			& \eqref{w1}, \eqref{w2}, \eqref{pk_fdma},
		\end{align}
	\end{subequations}
	where $R_k = w_k \log_2 \left(1+\frac{p_k\left|\mathbf{h}^H_{\mathrm{I},k} \left(\mathbf{x}_{\text{p}, k}\right) \mathbf{g} \left(\mathbf{x}_{\text{p}, k}\right)\right|^2}{N w_k \sigma^2_{\mathrm{I},k}}\right)$.
	The subproblem can be verified as a convex problem, which can be solved via standard solvers like CVX~\cite{grant2014cvx} to obtain the solutions of $\mathbf{w}$ and $\mathbf{p}$.
	
	The computation complexities of the PSO algorithm and resource allocation solution are $\mathcal{O}\left(L_{\max} |\mathcal{S}| N\right)$ and $\mathcal{O}\left(K^{3.5}\right)$, respectively, where $L_{\max}$ is the number of iteration for PSO, and $|\mathcal{S}|$ is the size of the particle swarm.
	Therefore, the overall computation complexity is $\mathcal{O}\left(\frac{1}{\varepsilon_3}\left(L_{\max} |\mathcal{S}| N + K^{3.5}\right)\right)$, where $\varepsilon_3$ is the convergence threshold for FDMA.
	
	\subsubsection{TDMA-Assisted PASS}
	
	Again, we employ the $\epsilon$-constraint method to transform the MOOP \eqref{moop_tdma} into an SOOP, which can be expressed as
	\begin{subequations}
		\label{tdma_soop}
		\begin{align}
			& \max_{\mathbf{X}_\text{p}, \boldsymbol{\tau}}  \min_{k \in \mathcal{K}}  R_{k}, \\
			\rm{s.t.} \ \
			& E_j \geq \epsilon, \forall j \in \mathcal{J}, \label{E_j3}\\
			& \eqref{spacing2}, \eqref{range2}, \eqref{tau1}, \eqref{tau2}.
		\end{align}
	\end{subequations}	
	Similarly to that for FDMA, we employ the AO algorithm again for itaratvely optimizing $\mathbf{X}_\text{p}$ and $\boldsymbol{\tau}$.
	
	\textbf{Optimizing $\{\mathbf{X}_\text{p}\}$}:
	Unlike FDMA and NOMA, TDMA offers the flexibility to dynamically optimize the PAs positions independently in each time slot for simultaneously maximizing the data rate of the served IU and providing energy harvesting for all EUs.
	For each time slot, the pinching beamforming subproblem can be expressed as
	\begin{subequations}
		\label{tdma_x}
		\begin{align}
			& \max_{\mathbf{x}_{\text{p},k}, k \in \mathcal{K}}  R_k, \\
			\rm{s.t.} \ \
			& E_j^k \geq \epsilon - \sum_{i=1, i \neq k}^{K} E_j^i, \forall k \in \mathcal{K}, \forall j \in \mathcal{J}, \label{tdma_ej} \\
			& \eqref{spacing1}, \eqref{range1},
		\end{align}
	\end{subequations}
	where the constraint \eqref{tdma_ej} ensures the minimum energy harvested requirement.
	Similar to the PAs positions optimization problem in the FDMA case, this subproblem can be solved by the PSO algorithm as given in \textbf{Algorithm~\ref{solve_x_p}}.
	In each iteration of the AO algorithm, the antenna position matrix $\mathbf{X}_{\text{p}}$ is updated by independently optimizing the PA positions within each time slot.
	
	\textbf{Optimizing $\{\boldsymbol{\tau}\}$}:
	Given the PA position matrix $\mathbf{X}_{\text{p}}$, the remaining problem is to optimize the time allocation vector, which is formulated as
	\begin{subequations}
		\begin{align}
			& \max_{\boldsymbol{\tau}}  \min_{k \in \mathcal{K}}  R_{k}, \\
			\rm{s.t.} \ \
			& \eqref{tau1}, \eqref{tau2}, \eqref{E_j3}. 
		\end{align}
	\end{subequations}
	
	By introducing the auxiliary variable $\xi$ that satisfies $R_k \geq \xi, \forall k \in \mathcal{K}$, the subproblem can be recast as
	\begin{subequations}
		\label{tdma_t}
		\begin{align}
			& \max_{\boldsymbol{\tau}, \xi}  \xi, \\
			\rm{s.t.} \ \
			& \eqref{tau1}, \eqref{tau2}, \eqref{E_j3}, \\
			& \xi - R_k \leq 0, \forall k \in \mathcal{K},
		\end{align}
	\end{subequations}
	where $R_k = \tau_k \log_2 \left(1+\frac{P}{N \sigma^2_{\mathrm{I},k}} \left|\mathbf{h}^H_{\mathrm{I},k} \left(\mathbf{x}_{\text{p}, k}\right) \mathbf{g} \left(\mathbf{x}_{\text{p}, k}\right)\right|^2\right)$.
	It is observed that \eqref{tdma_t} is a convex problem, which can be solved by existing solvers in CVX~\cite{grant2014cvx}.
	
	The overall AO algorithm for TDMA is given in \textbf{Algorithm~\ref{tdma_solution}}.
	The computation complexity of the pinching beamforming for one time slot is $\mathcal{O}\left(L_{\text{max}} |\mathcal{S}| N \right)$.
	As this process is repeated for the $K$ time slots, the computational complexity of pinching beamforming subproblem is $\mathcal{O}\left(L_{\text{max}} |\mathcal{S}| N K \right)$.
	And the computational complexity of the time allocation subproblem is $\mathcal{O}\left(K^{3.5}\right)$.
	Therefore, the overall complexity is $\mathcal{O}\left(\frac{1}{\varepsilon_4}\left(K^{3.5} + L_{\text{max}} |\mathcal{S}| N K \right)\right)$, where $\varepsilon_4$ is the convergence threshold for TDMA.
	
	\begin{algorithm}[t]
		\caption{Proposed Overall PSO-Based AO Algorithm to Solve Problem \eqref{tdma_soop}}
		\label{tdma_solution}
		\begin{algorithmic}[1]
			\STATE Initialize the PA position matrix $\mathbf{X}_\text{p}$ and time allocation vector $\boldsymbol{\tau}$. Set convergence criterion $\varepsilon_4$.
			\REPEAT
			\STATE Set the iteration index $k = 1$.
			\REPEAT
			\STATE Given $\{\boldsymbol{\tau}, \mathbf{x}_{\text{p}, i}, \forall i \in \mathcal{K}, i \neq k\}$, update $\mathbf{x}_{\text{p}, k}$ by solving the problem \eqref{tdma_x}.
			\STATE $k = k + 1$.
			\UNTIL $k=K+1$.
			\STATE Given $\mathbf{X}_\text{p}$, update $\boldsymbol{\tau}$ by solving the problem \eqref{tdma_t}.
			\UNTIL{the fractional increment of the objective function is below $\varepsilon_4$.}
		\end{algorithmic}
	\end{algorithm}

	\subsubsection{NOMA-Assisted PASS} 
	
	With the $\epsilon$-constraint method, the MOOP for FDMA can be transformed to the following SOOP:
	\begin{subequations}
		\label{noma_soop}
		\begin{align}
			& \max_{\mathbf{x}_\text{p}, \mathbf{a}}  \min_{k \in \mathcal{K}}  R_k, \\
			\rm{s.t.} \ \
			& E_j \geq \epsilon, \forall j \in \mathcal{J}, \label{e_j} \\
			& \eqref{spacing1}, \eqref{range1}, \eqref{gain}, \eqref{alpha1}, \eqref{alpha2}.
		\end{align}
	\end{subequations} 
	Considering the non-convexity of problem \eqref{noma_soop} and the mutual coupling between optimization variables, we employ the AO algorithm again to iteratively optimize $\mathbf{x}_{\text{p}}$ and $\mathbf{a}$.
	
	\textbf{Optimizing $\{\mathbf{x}_\text{p}\}$}: With given power allocation coefficient vector $\mathbf{a}$, the pinching beamforming optimization subproblem can be expressed as
	\begin{subequations}
		\label{r_k}
		\begin{align}
			& \max_{\mathbf{x}_\text{p}}  \min_{k \in \mathcal{K}}  R_{k}, \\
			\rm{s.t.} \ \
			& \eqref{spacing1}, \eqref{range1}, \eqref{gain}, \eqref{e_j}.
		\end{align}
	\end{subequations} 
	Note that, compared to the cases of FDMA and TDMA, the main challenge for solving the pinching beamforming problem under the NOMA protocol relies on the following aspect. 
	Since the channel gains are greatly affected by the pinching beamforming, the pinching beamforming design needs to take the decoding order into consideration. 
	Specifically, for any users $k$ and $k'$, $\forall k, k' \in \mathcal{K}$, if $\left|\mathbf{h}^H_{\mathrm{I},k} \left(\mathbf{x}_\text{p}\right) \mathbf{g} \left(\mathbf{x}_\text{p}\right)\right|^2 \geq \left|\mathbf{h}^H_{\mathrm{I},k'} \left(\mathbf{x}_\text{p}\right) \mathbf{g} \left(\mathbf{x}_\text{p}\right)\right|^2$, we have $\pi(k) > \pi(k')$.
	Therefore, by invoking the PSO algorithm proposed for the FDMA case as shwon in Algorithm 2, the algorithm designated for the NOMA case should involve exhaustively searching over all possible decoding order combinations for determining the optimal one in each PSO iteration.
	
	\textbf{Optimizing $\{\mathbf{a}\}$}:
	Next, we focus on the optimization of the power allocation coefficient vector $\mathbf{a}$ with the given PAs positions $\mathbf{x}_\text{p}$ and decoding order $\boldsymbol{\pi}$.
	The power allocation subproblem is given by
	\begin{subequations}
		\begin{align}
			& \max_{\mathbf{a}}  \min_{k \in \mathcal{K}}  R_{k}, \\
			\rm{s.t.} \ \
			& \eqref{alpha1}, \eqref{alpha2}.
		\end{align}
	\end{subequations}
	By introducing the auxiliary variable $\xi$, which satisfies $R_k \geq \xi, \forall k \in \mathcal{K}$, the subproblem can be recast by
	\begin{subequations}
		\label{org_xi}
		\begin{align}
			& \max_{\mathbf{a}, \xi}  \xi, \\
			\rm{s.t.} \ \
			& R_k \geq \xi, \forall k \in \mathcal{K}, \label{rk} \\
			& \eqref{alpha1}, \eqref{alpha2}.
		\end{align}
	\end{subequations} 
	Since constraint \eqref{rk} remains non-convex, we invoke the SCA method to deal with it.
	To facilitate the application of SCA, we assume that the decoding order is $ \pi(k) = k$, and we introduce auxiliary variables $\mathbf{b} = \left[b_1, ..., b_{K+1}\right]$, where $b_k = \sum^{K}_{i=k} \alpha_i \frac{P}{N}$ and $b_{K+1} \triangleq 0$.
	Thus, the achievable rate of the $k$-th IU can be expressed as
	\begin{equation}
		\begin{aligned}
			R_k & = \log_2 \left(1+\frac{\alpha_k \frac{P}{N} \left|\mathbf{h}^H_{\mathrm{I},k} \left(\mathbf{x}_\text{p}\right) \mathbf{g} \left(\mathbf{x}_\text{p}\right)\right|^2}{\sum^{K}_{i=k+1} \alpha_i \frac{P}{N} \left|\mathbf{h}^H_{\mathrm{I},k} \left(\mathbf{x}_\text{p}\right) \mathbf{g} \left(\mathbf{x}_\text{p}\right)\right|^2 +\sigma^2_{\mathrm{I},k}}\right)\\
			& = \log_2 \left(b_k \left|\mathbf{h}^H_{\mathrm{I},k} \left(\mathbf{x}_\text{p}\right) \mathbf{g} \left(\mathbf{x}_\text{p}\right)\right|^2 +\sigma^2_{\mathrm{I},k}\right) \\
			& \quad - \log_2\left(b_{k+1} \left|\mathbf{h}^H_{\mathrm{I},k} \left(\mathbf{x}_\text{p}\right) \mathbf{g} \left(\mathbf{x}_\text{p}\right)\right|^2 +\sigma^2_{\mathrm{I},k}\right).
		\end{aligned}
	\end{equation}
	Observing that $R_k$ is the difference of two concave functions.
	In the $l$-th SCA iteration, the concave lower bound of the achievable rate can be obtained by applying the first-order Taylor expansion to the second term at the given local point $b^{(l)}_{k+1}$.
	The expansion of the achievable rate is expressed as
	\begin{equation}
		\begin{aligned}
			R_k \geq \overline{R}_{k}& = \log_2 \left(b_k \left|\mathbf{h}^H_{\mathrm{I},k} \left(\mathbf{x}_\text{p}\right) \mathbf{g} \left(\mathbf{x}_\text{p}\right)\right|^2 +\sigma^2_{\mathrm{I},k} \right) \\
			& \quad - \log_2 \left(b^{(l)}_{k+1} \left|\mathbf{h}^H_{\mathrm{I},k} \left(\mathbf{x}_\text{p}\right) \mathbf{g} \left(\mathbf{x}_\text{p}\right)\right|^2 +\sigma^2_{\mathrm{I},k}\right) \\
			& \quad - \frac{\left|\mathbf{h}^H_{\mathrm{I},k} \left(\mathbf{x}_\text{p}\right) \mathbf{g} \left(\mathbf{x}_\text{p}\right)\right|^2 \left(b_{k+1} - b^{(l)}_{k+1}\right) \log_2 e}{b^{(l)}_{k+1} \left|\mathbf{h}^H_{\mathrm{I},k} \left(\mathbf{x}_\text{p}\right) \mathbf{g} \left(\mathbf{x}_\text{p}\right)\right|^2 +\sigma^2_{\mathrm{I},k}}.
		\end{aligned}
	\end{equation}
	By substituting the concave lower bound of the achievable rate $\overline{R}_{k}$ into constraint \eqref{rk}, the original problem can be transformed into the following convex one:
	\begin{subequations}
		\label{sub_b_xi}
		\begin{align}
			& \max_{\mathbf{b}, \xi}  \xi, \\
			\rm{s.t.} \ \
			& b_1 \leq \frac{P}{N}, \\
			& b_1 - b_2 \geq b_2 - b_3 \geq ... \geq b_K \geq 0, \\
			& \xi - \overline{R}_{k} \leq 0, \forall k \in \mathcal{K}, \label{xi_R}
		\end{align}
	\end{subequations} 
	The problem \eqref{sub_b_xi} can be solved directly by existing convex solver like CVX~\cite{grant2014cvx}.
	Furthermore, the power allocation solutions are given by the following expression:
	\begin{equation}
		\alpha^{*}_k = \frac{N}{P} (b^{*}_k - b^{*}_{k+1}), \forall k \in \mathcal{K},
	\end{equation}
	where $b^{*}_k$ and $b^{*}_{k+1}$ are the optimal solutions obtained by solving the problem \eqref{sub_b_xi}.
	
	\begin{algorithm}[t]
		\caption{Proposed Overall PSO-Based AO Algorithm to Solve Problem \eqref{noma_soop}}
		\label{solve_x_a}
		\begin{algorithmic}[1]
			\STATE Initialize PAs positions $\mathbf{x}_\text{p}$, and power allocation vector $\mathbf{a}$. Set convergence criterion $\varepsilon_5$.
			\REPEAT
			\STATE Given $\mathbf{a}$, update $\boldsymbol{\pi}$ by exhaustively searching and update $\mathbf{x}_\text{p} $ by solving problem \eqref{r_k} with \textbf{Algorithm~\ref{solve_x_p}}.
			\STATE Given $\mathbf{x}_\text{p}$, update $\mathbf{a}$ by solving \eqref{org_xi} with the SCA method.
			\UNTIL{the fractional increment of the objective function is below $\varepsilon_5$.}
		\end{algorithmic}
	\end{algorithm}
	The overall procedure for solving problem \eqref{noma_soop} is summarized in \textbf{Algorithm~\ref{solve_x_a}}. 
	The computational complexity of the PSO algorithm is $\mathcal{O}\left(L_{\max} |\mathcal{S}| N K!\right)$, where $K!$ is the all possible decoding order combinations.
	The computational complexity of power allocation optimization is $\mathcal{O}\left(I_{\text{iter}} K^{3.5}\right)$, where $I_{\text{iter}}$ is the number of iterations for SCA.
	Thus, the overall computation complexity is $\mathcal{O}\left(\frac{1}{\varepsilon_5}\left(L_{\max} |\mathcal{S}| N K! + I_{\text{iter}} K^{3.5}\right)\right)$, where $\varepsilon_5$ is the convergence threshold for NOMA.

\section{Simulation Results}

In this section, the simulation results are provided to validate the effectiveness of the PASS-assisted SWIPT system.

\subsection{Simulation Setup}

Unless otherwise specified, the simulations adopt the following parameter setups.
The carrier frequency is set to $f_c = 28$ GHz, the noise power for each user is $\sigma^2_{u} = -90$ dBm, the energy conversion efficiency for each EU is $\zeta = 50\%$, and the minimum spacing between PAs is set to $\Delta = \frac{\lambda}{2}$ to avoid the coupling effect. 
For the waveguide, the height is set to $d = 3$ meter, the maximum length is $L = 20$ meter, and the effective refractive index is $n_\text{eff} = 1.4$.
Moreover, for the PSO algorithm, the particle swarm size is set to $|\mathcal{S}| = 300$, $\omega_\text{max} = 0.9$, $\omega_\text{min} = 0.1$, $c_1 = c_2 = 1.5$, $\rho_1 = \rho_2 = \rho_3 = 10^8$, and $L_\text{max} = 300$.
Convergence thresholds are set to $\varepsilon_1 = \varepsilon_2 = \varepsilon_3 = \varepsilon_4 = \varepsilon_5 = 1 \times 10^{-3}$.

\subsection{Single IU/EU Case}

\begin{figure}
	\centering
	\begin{subfigure}{\linewidth}
		\centering
		\includegraphics[width=0.9\linewidth]{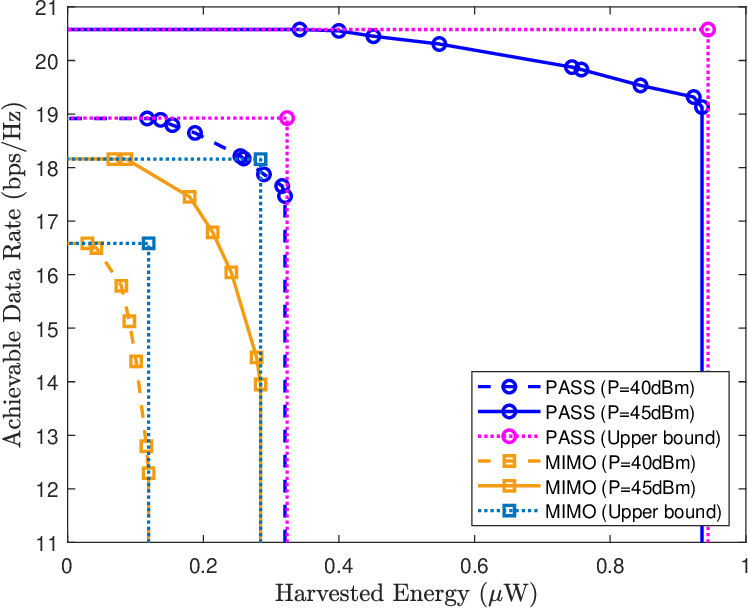}
		\caption{Rate-energy region with different transmit power.}
		\label{one_user_P}
	\end{subfigure}
	\begin{subfigure}{\linewidth}
		\centering
		\includegraphics[width=0.9\linewidth]{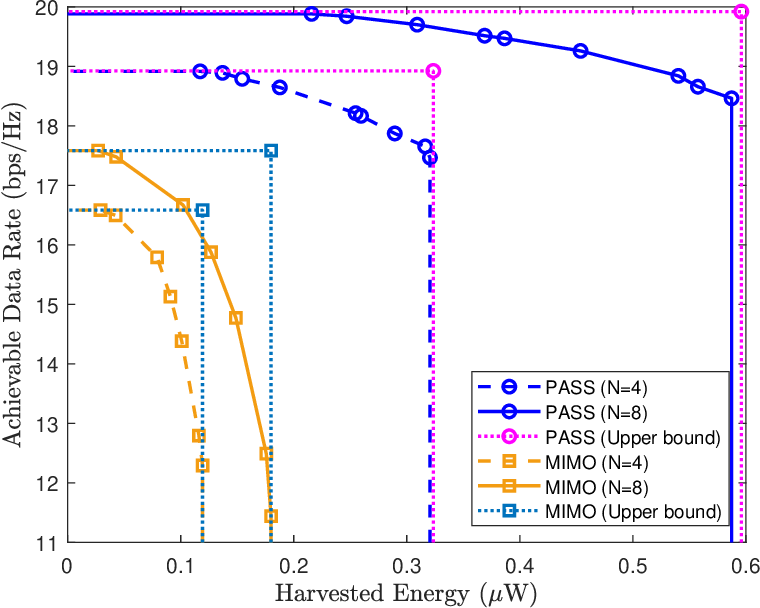}
		\caption{Rate-energy region with different number of PAs.}
		\label{one_user_N}
	\end{subfigure}
	\caption{Achievable rate-energy region for the single IU/EU scenario.}
	\label{fig:region_single}
\end{figure}

In this subsection, we simulate the rate-energy region achieved by the proposed two-stage pinching beamforming algorithm for the single IU/EU scenario.
For performance comparison, we consider the conventional MIMO system as the benchmark, where a BS equipped with one RF chain and the same number of antennas as PASS is deployed at the location $[-\frac{L}{2}, 0, d]^T$, with an antenna spacing of half the wavelength.

In Fig.~\ref{fig:region_single}, we demonstrate the achievable rate-energy region for the single IU/EU case.
As illustrated in Fig.~\ref{fig:region_single} (a) and Fig.~\ref{fig:region_single} (b), the achievable rate-energy region of PASS expands as the transmit power and the number of PAs increase.
This is because, on the one hand, higher transmit power brings in improved gain at both the IU and EU. 
On the other hand, a higher number of PAs leads to more spatial flexibility for enhancing the pinching beamforming performance.
Moreover, as $\rho$ gradually increases, the achievable rate decreases while the harvested energy rises.
This is because a larger $\rho$ relaxes the achievable rate requirement, allowing the PAs to move along the waveguide towards EU to improve the energy harvesting efficiency.
For a comprehensive performance evaluation, we also provide the upper bounds for both the proposed PASS and conventional MIMO systems.
The upper bound for PASS is obtained by considering a single-user scenario (i.e., one IU or one EU) and applying the corresponding PA position optimization method as proposed in~\cite{XuRate2025}.
For the conventional MIMO system, its upper bound is achieved by the maximum ratio transmission approach in the same single-user scenario.
It is observed that PASS achieve a larger rate-energy region than the conventional MIMO system, benefiting from its ability to significantly reduce path loss by flexibly repositioning PAs closer to users.
It is also observed that the rate-energy region achieved by PASS is notably more rectangular than that achieved by the conventional MIMO system.   
In other words, a significant gap is also observed between the practical performance and the upper bound for the conventional MIMO system, while the practical performance of PASS is only slightly below its upper bound.
This is expected thanks to the more flexible adjustment of PAs positions that facilitates higher spatial DoFs, thus allowing for higher performance gain regarding to both communication rate and harvested energy.

\subsection{Multiple IUs and EUs Case}

This subsection presents the performance of PASS-assisted SWIPT system employing different multiple access schemes for the multiple IUs/EUs scenario. 
The number of IUs and EUs are set to $K=2$ and $J=2$, respectively.
To validate the credibility and effectiveness of the PASS-assisted SWIPT system, we consider the following benchmarks:

\begin{itemize}
	\item \textbf{Conventional single-antenna system} (labeled "\textbf{Con1}"): In this case, a single-antenna BS is deployed at the location $[-\frac{L}{2}, 0, d]^T$ and serves multiple IUs/EUs.
	The corresponding resource allocation optimization problem for each multiple access scheme is solved by applying the proposed methods detailed in Section III.
	\item \textbf{Conventional MIMO system} (labeled "\textbf{Con2}"):In this case, a BS is equipped with one RF chain and the same number of antennas as PASS.
	The antennas are centered at $[-\frac{L}{2}, 0, d]^T$ and spaced with half a wavelength.
	The received signal at the user is given by $y_{\text{u}} = \mathbf{h}^H_{\text{u}} \mathbf{w} s + n_{\text{u}}$, where $ \mathbf{h}_{\text{u}} \in \mathbb{C}^{N \times 1}$ and $\mathbf{w} \in \mathbb{C}^{N \times 1}$ are the complex channel and analog beamforming vectors, respectively.
	For each multiple access scheme, the corresponding analog beamforming and resource allocation subproblems are solved by applying the semi-definite relaxation~\cite{LiuSecrecy2014} and the methods detailed in Section III, respectively. 
	Notably, by exploiting the time-switching feature of TDMA, MIMO can generate distinct beams for different IUs in separate time slots.
\end{itemize}

\begin{figure}[t]
	\centering
	\includegraphics[width=3in]{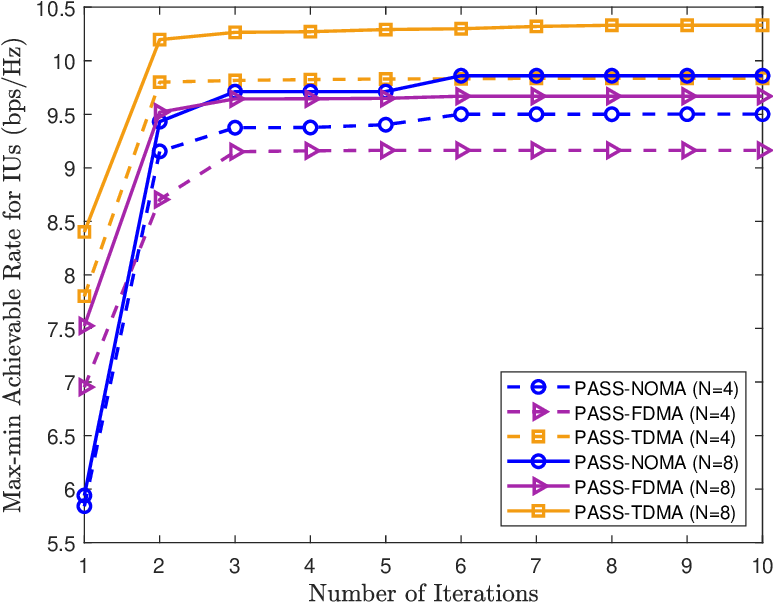}
	\caption{Convergence performance of the proposed PSO-baesd AO algorithms for the multiple IUs/UEs scenario, with $N=4$ and $N=8$.}
	\label{multi_user_iterations}
\end{figure}

In Fig.~\ref{multi_user_iterations}, we depict the convergence performance of the proposed PSO-based AO algorithms for different multiple access schemes, with a minimum harvested energy threshold of $0.05 \mu$W.
The results show that the max-min achievable rate increases rapidly at the beginning and converges within approximately 6 iterations for all schemes, which demonstrates the effectiveness of the proposed AO algorithm.
It is also observed that the max-min rate increases with the number of PAs. 
This is an expected result, as more PAs provide a higher pinching beamforming gain to the information signals directed at the IUs.

\begin{figure}[t]
	\centering
	\includegraphics[width=3in]{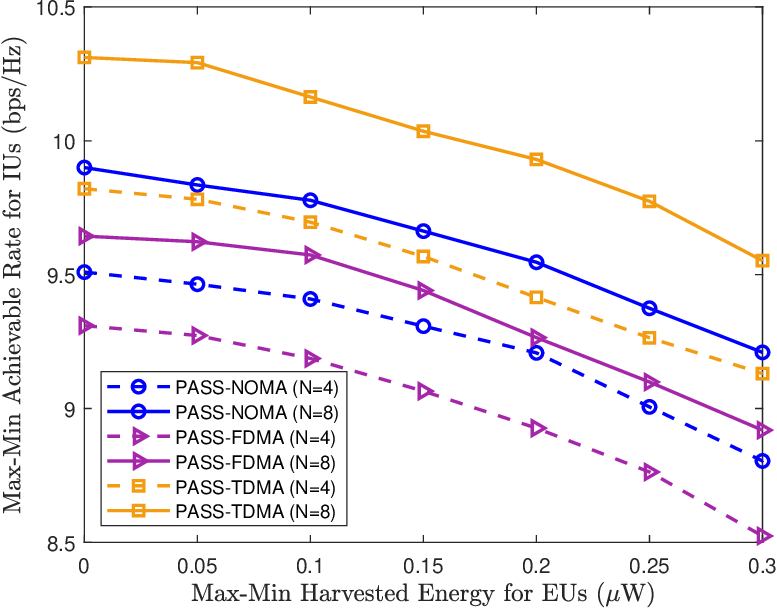}
	\caption{Achievable rate-energy region for the multiple IUs/UEs scenario, with $N = 4$ and $N = 8$.}
	\label{multi_user_tradeoff_4N_8N}
\end{figure}

In Fig.~\ref{multi_user_tradeoff_4N_8N}, we explore the achievable rate-energy region for the multiple IUs/EUs scenario with $N=4$ and $N=8$.
A fundamental tradeoff is observed across all multiple access schemes, where the max-min achievable rate decreases as the max-min harvested energy by EUs increases. 
The tradeoff stems from the inherent resource competition, as optimizing PAs positions and resource allocation to enhance energy harvested by EUs inevitably compromises the achievable rate for IUs.
Moreover, under identical conditions, the max-min achievable rate for IUs of TDMA is higher than that of NOMA and FDMA.
This is because, the combination of TDMA and PASS exhibits a unique advantage, as it can utilize the extra time-domain flexibility for adjusting PAs positions.
By redeploying PAs within each time slot, TDMA gains additional DoF, enabling it to satisfy the harvested energy constraints and maximize the achievable rates.
Furthermore, NOMA outperforms FDMA by serving all users simultaneously within the same time-frequency resource block.

\begin{figure}[t]
	\centering
	\includegraphics[width=3in]{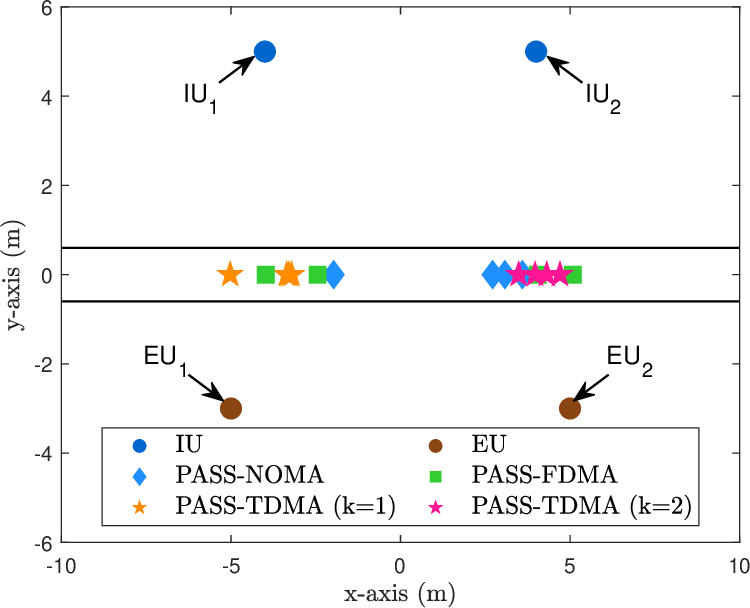}
	\caption{PA deployment locations for different multiple access schemes.} 
	\label{PAs_positions}
\end{figure}

In Fig.~\ref{PAs_positions}, we provide the PAs deployment positions for the multiple IUs/EUs scenario with IU$_1$ at $[-4, 5, 0]$ meter, IU$_2$ at $[4, 5, 0]$ meter, EU$_1$ at $[-5, -3, 0]$ meter, and EU$_2$ at $[5, -3, 0]$ meter, with $N = 4$.
For FDMA, it is interesting to find that, PAs distribution is symmetrical, i.e., two PAs are placed near IU$_1$/EU$_1$, and the other two are placed near IU$_2$/EU$_2$. 
This is expected, as all the PAs need to serve all users simultaneously, and thus, PAs deployment need to consider the average distance to all users.
However, the deployment strategy becomes different for the NOMA case, with three PAs deployed on the waveguide near IU$_2$, while only one is located near IU$_1$. 
This deployment strategy is designed to enlarge the channel gain disparity among different users, which is a unique and advantageous configuration for NOMA.
Consequently, the signal intended for IU$_1$ is allocated more transmit power and decoded first.
Specifically, for TDMA, at the first time slot, i.e., $k=1$, when the IU$_1$ is served, all PAs are deployed near IU$_1$, whereas at the second time slot, i.e., $k=2$, when IU$_2$ is served, PAs are deployed near IU$_2$. 
Benefiting from this time switching flexibility of PASS, users with TDMA can achieve higher channel gain to enhance the performance of SWIPT system.
Consequently, this flexible deployment strategy enhances the channel gain for each user during their allocated time slot, significantly improving SWIPT system performance.

\begin{figure}[t]
	\centering
	\includegraphics[width=3in]{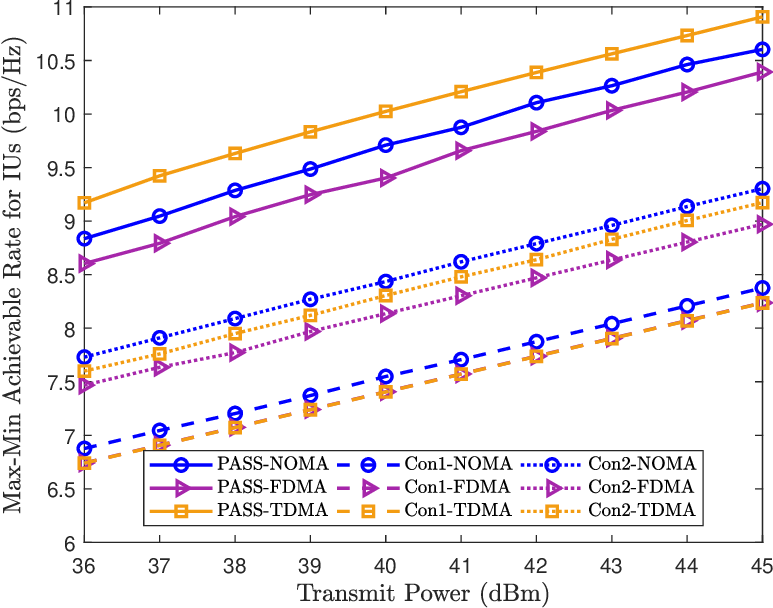}
	\caption{Max-min achievable rate versus the transmit power for the multiple IUs/UEs scenario, with $N = 4$ and $\epsilon = 0.05 \mu$W.}
	\label{multi_user_P}
\end{figure}

In Fig.~\ref{multi_user_P}, we present the max-min achievable rates versus the transmit power for the multiple IUs/EUs scenario, with the minimum harvested energy threshold $0.05 \mu$W.
As expected, the PASS-assisted SWIPT system can achieve higher max-min achievable rates than both the conventional single antenna and MIMO systems.
This is because, PASS can flexibly move PAs towards users to establish stronger LoS links, which leads to lower path loss.
Moreover, by exploiting the time-switching feature of TDMA, the conventional MIMO system can generate distinct analog beams for the served users in different time slots, thus achieving a higher max-min achievable rate than that of FDMA.
In contrast to PASS, the max-min achievable rate for MIMO of TDMA is lower than that of NOMA.
This observation suggests that TDMA provides greater performance gain for the PASS-assisted system thanks to more flexible PAs positions adjustment across different time slots.

\begin{figure}[t]
	\centering
	\includegraphics[width=3in]{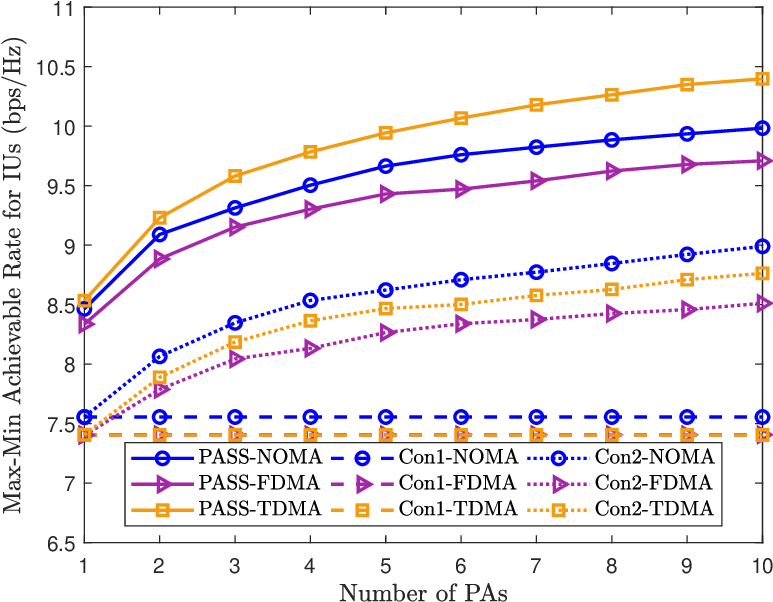}
	\caption{Max-min achievable rate versus the number of PAs for the multiple IUs/UEs scenario, with $P = 40$ dBm and $\epsilon = 0.05 \mu$W.}
	\label{multi_user_N}
\end{figure}

In Fig.~\ref{multi_user_N}, we further investigate the max-min achievable rates for the multiple IUs/EUs scenario, considering different numbers of PAs and a minimum harvested energy threshold of $0.05 \mu$W.
It is first observed that the max-min achievable rate improves for both PASS and conventional MIMO systems as the number of antennas increases. 
This improvement stems from the higher beamforming gains and additional spatial DoF afforded by a larger number of antennas, which enhance both information transmission and energy harvesting.
Furthermore, a key distinction from the conventional MIMO system is that the performance gain of PASS with TDMA over NOMA and FDMA becomes more significant with a larger number of PAs.
This is because the benefits of antenna deployment flexibility are improved as the number of PAs increases.
However, this enhanced flexibility introduces a higher cost in computational complexity and greater deployment challenges.

\section{Conclusion}

This paper investigated a PASS-assisted SWIPT system, considering both the single IU/EU and multiple IUs/EUs scenarios.
For the single IU/EU scenario, a weighted sum optimization problem was formulated to explore the rate-energy region.
To solve the non-convex problem, a two-stage solution was proposed for PA position optimization to achieve the trade-off between data rate and harvested energy.
Then, for the multiple IUs/EUs scenario, we formulated the MOOPs that simultaneously maximize the minimum data rate and minimum harvested energy, under three different multiple access schemes, i.e., FDMA, TDMA, and NOMA.
By employing the $\epsilon$-constraint method, each MOOP was transformed into the corresponding SOOP.
To solve these problems, the PSO-based AO algorithms were proposed to jointly optimize the PAs positions and resource allocation alternately.
Simulation results showed that the PASS-assisted SWIPT system could achieve larger rate-energy region compared to the conventional antenna systems.
Moreover, owing to the deployment flexibility of PASS, the PASS-assisted SWIPT system with TDMA can achieve a better performance than NOMA and FDMA.

\bibliographystyle{IEEEtran}
\bibliography{Ref}

\end{document}